\documentclass[twocolumn]{aastex62}
\usepackage{appendix,hyperref}

\usepackage[T1]{fontenc}
\usepackage{outlines}

\DeclareRobustCommand{\VAN}[3]{#2}
\let\VANthebibliography\thebibliography
\def\thebibliography{\DeclareRobustCommand{\VAN}[3]{##3}\VANthebibliography}

\hypersetup{linkcolor=blue,citecolor=blue,filecolor=cyan,urlcolor=magenta}
\usepackage{graphicx}   
\usepackage{amsmath}    
\usepackage{amssymb}    
\usepackage{newtxtext,newtxmath}
\usepackage{color}
\usepackage{threeparttable}
\usepackage{hyperref}
\usepackage{ulem}

\def\MBH{M_{_{\rm BH}}}
\def\mbh6{m_{_{\rm BH,6}}}
\def\Ms{M_{\star}}
\def\ms{m_{\star}}

\def\rg{r_{\rm g}}

\newcommand{\lta}{\lower 2pt \hbox{$\, \buildrel {\scriptstyle <}\over {\scriptstyle \sim}\,$}}
\newcommand{\gta}{\lower 2pt \hbox{$\, \buildrel {\scriptstyle >}\over {\scriptstyle \sim}\,$}}

\begin{document}

\title{ Follow the Mass - A Concordance Picture of Tidal Disruption Events}
\shorttitle{Follow the Mass}

\author[0000-0002-2995-7717]{Julian Krolik}
\affiliation{Physics and Astronomy Department, Johns Hopkins University, Baltimore, MD 21218, USA}
\correspondingauthor{Julian Krolik}
\email{jhk@jhu.edu}

\author[0000-0002-7964-5420]{Tsvi Piran}
\affiliation{Racah Institute for Physics, The Hebrew University, Jerusalem, 91904, Israel}

\author[0000-0003-2012-5217]{Taeho Ryu}
\affiliation{Max-Planck-Institut f\"ur Astrophysik, Karl-Schwarzschild-Str. 1, Garching, 85748, Germany}
\affiliation{JILA, University of Colorado and National Institute of Standards and Technology, 440 UCB, Boulder, 80308 CO, USA}
\affiliation{Department of Astrophysical and Planetary Sciences, 391 UCB, Boulder, 80309 CO, USA}


\begin{abstract}
Three recent global simulations of tidal disruption events (TDEs) have produced, using  different numerical techniques and parameters, very similar pictures of their dynamics. In typical TDEs, after the star is disrupted by a supermassive black hole, the bound portion of the stellar debris follows highly eccentric trajectories, reaching apocenters of several thousand gravitational radii.  Only a very small fraction is captured upon returning to the vicinity of the supermassive black hole.  Nearly all the debris returns to the apocenter, where shocks produce a thick irregular cloud on this radial scale and power the optical/UV flare. These simulation results imply that over a few years, the thick cloud settles into  an accretion flow responsible for the long term emission. Despite not being designed to match observations, and without adjusting any parameters, the dynamical picture given by the three simulations aligns well with observations of typical events, correctly predicting the flares' typical total radiated energy, luminosity, temperature and emission line width.  On the basis of these predictions, we provide an updated method ({\sc TDEmass}) to infer the stellar and black hole masses from a flare's peak luminosity and temperature.  This picture also correctly predicts that the luminosity observed years after the flare should be nearly constant.  In addition, we show that in a magnitude-limited survey, if the intrinsic rate of TDEs is independent of black hole mass, the detected events will preferentially have black hole masses $\sim 10^{6.3 \pm 0.3} M_\odot$ and stellar masses $\sim 1 M_\odot$, with the width of the mass distribution for disrupted stars sensitive to the stellar mass function in the host galaxy's center.
\end{abstract}

\keywords{}

\section{Introduction}
\label{sec:intro}

Tidal disruption events (TDEs) are inherently dramatic: a star is ripped apart by the gravity of a supermassive black hole. The result is a flare in which, for a month or two, the former star radiates with $\sim 10^{10}\times$ its ordinary luminosity. Over the duration of the flare (typically months), TDEs are among the most luminous transients known.  Once hard to find, there are now $\sim 100$ known examples \citep{SDSS,ASASSN,PANSTARRS,ZTF,Hammerstein+2023}, and many more can be expected from new instruments (e.g., the Rubin Observatory: \citealt{BricmanGomboc2020,Szekerczes+2024} and ULTRASAT: \citealt{Ben-Ami+2022}) due to go online soon.  

As TDEs involve infall of matter onto a black hole, they are of significant interest as a tool to explore dynamical accretion processes.  They also have potential interest as a probe of the black hole population in galactic nuclei and the dynamical relationship between those nuclear black holes and the surrounding stars. 

In addition to these reasons, they also merit attention as truly multiwavelength systems \citep[see][for the most recent review]{WeversRyu2023}. TDEs have been observed mostly in the optical, but at times, they shine in X-rays and in radio as well. The emission in different bands typically comes at different times and emerges from different locations, indicating that multiple mechanisms contribute to their radiation processes.

To identify where and how the radiation is emitted, the first task is to track where the debris mass goes.   When the star's orbit is effectively parabolic and the star has been fully disrupted, half the star's mass is unbound and escapes to infinity, while the other half remains bound to the black hole \citep{Rees1988}. Although there is a firm consensus that the bound mass is initially placed on highly-eccentric orbits
whose apocenters are thousands of gravitational radii, $\sim O(10^2) \times$ the stellar pericenter, hitherto there has been little agreement about its whereabouts following its first return to the pericenter. 
The oldest, and still most prevalent, view has been that the bound material immediately forms a compact accretion disk on the pericenter scale \cite{Rees1988}. This picture predicts a very luminous X-ray source, but the observed spectra have luminosities $\sim 10^{-2}\times$ this prediction, and shapes well fit by black bodies with temperatures $\gtrsim 10^4$~K and radiating areas corresponding to the apocenter scale (see \citet{vanVelzen2020SSRV,Hammerstein+2023} for two recent data compilations). To explain these unexpected observations, many (e.g., \citet{Strubbe2009,SQ2011, MetzgerStone2016}) have suggested that the X-rays are reprocessed by different material associated with the event whose photosphere happens to be at this distance from the black hole.
Alternatively, it has also been suggested \citep{Shiokawa+2015} that, rather than forming a compact disk, the bound material returns to the apocenter region and, upon colliding with later-returning bound debris, forms a large irregular cloud on a scale similar to the apocenter distance.  In this picture \citep{Piran+2015}, the shocks associated with these collisions power the flare.

The question of where the debris finds itself shortly after the disruption is given further importance by the fact that the energy available for radiation is exactly the binding energy of the debris orbits.  The binding energy per unit mass is, of course, inversely proportional to the semimajor axis of a given fluid element's orbit.  Given the large discrepancy between the ratio of radiated energy to debris mass and the energy per unit mass released in relativistic accretion, the orbital distribution is, therefore, of prime dynamical interest.

Large-scale numerical simulations treating most of the bound debris have the power to resolve these questions, but hitherto their computational expense has been prohibitive due to the extremely large dynamical range in lengthscales inherent to the problem.  Consequently, early global simulations were run with parameter choices that, although unlikely to describe real events, made the runs feasible \citep{Rosswog2008,Shiokawa+2015}. In these simulations, nearly all the bound matter passes again through the apocenter region after its first return to the vicinity of the black hole. However, as the simulation parameters were not realistic, these results have not been widely accepted as relevant to observed TDEs.

Recently, the numerical barriers have been overcome by three different groups \citep{Ryu2023b,SteinbergStone2024,Price+2024} using very different numerical methods, but all of them employing well-understood physics. In this paper, we will show that these simulations largely agree on where the bound debris goes after its first return to the vicinity of the black hole: $>99\%$ is placed on orbits extending to thousands of gravitational radii, while $<1\%$ is confined within distances comparable to the disrupted star's pericenter.

In addition, we will outline the striking observational implications stemming from this consensus. Without the use of any adjustable parameters, they are in good agreement with typical observed values of 
the luminosity and spectral shape at the peak of the flare, the total radiated energy during the flare peak, the character of the long-term lightcurve, and the stellar and black hole masses most commonly inferred from detected events.

However, it is important to note that this model, although applicable to the majority of the TDE population, does not apply when the black hole mass $M_{\rm BH} \gtrsim 10^7 M_\odot$.  For these more massive black holes, the maximum stellar pericenter at which a total disruption takes place is $\lesssim 10 r_g$ ($r_g \equiv GM_{\rm BH}/c^2$) \citep{Ryu+2020d}.  In such cases, relativistic apsidal precession is so strong that it qualitatively changes the characteristic behavior of the debris.

We will first briefly summarize the relevant observational results, focusing on TDEs of the most commonly seen variety, those dominated by an optical/UV flare (Sec.~\ref{sec:briefobsns}).  Next (in Sec.~\ref{sec:quants}) we will remind readers of the tidal debris' characteristic lengthscales, orbital energies, etc.  Following these presentations of background, we will summarize the three new global simulations (Sec.~\ref{sec:simulations}).  The heart of our work will appear in Sec.~\ref{sec:implications}, where we demonstrate how the shared results of the simulations lead to important statements about both the underlying dynamics of TDEs and many of their observed properties.  Our conclusions will then be summarized in Sec.~\ref{sec:conclusions}.

\section{A brief summary of observations }
\label{sec:briefobsns}

\subsection{The optical flare}

In most optical/UV TDEs, the optical/UV flux rises quickly (over a few weeks) to a peak of $\sim 3 \times 10^{43}-3 \times 10^{44}$ erg~s$^{-1}$, followed by a lengthy (a few months) decline.  The spectra are well-fit by single-temperature Planck functions, whose
typical temperatures are $\sim 3 \times 10^4$ K and corresponding black-body emitting radii are  $10^{14-15}$ cm \citep{vanVelzen2020SSRV,Hammerstein+2023}. When lines are observed, their typical width corresponds to $\sim 5000$ km/sec \citep[see e.g., for a review][]{Gezari2021}.
For a black hole mass $\sim 10^6 M_\odot$, the scale of the emitting area is consistent with the scale of orbits with the speed inferred from the lines.

In some cases, the luminosity declines from its peak $\propto t^{-\alpha}$ with $\alpha \simeq 5/3$, the expected slope of the mass infall rate's decline \citep{Rees1988,Phinney1989}.  However, the entire sample exhibits a wider range of lightcurve power-law indices, $1 \lesssim \alpha \lesssim 3$ \citep{Hammerstein+2023}. Approximate integration of lightcurves yields a total radiated energy $\sim 10^{50.5 \pm 0.5}$~erg (the $\pm 0.5$ in the exponent refers to the standard deviation in the logarithm). At late times, years or more after the peak, the decay is much shallower.  For those events observed $\gtrsim 0.5$~yr past the peak, although there is a good deal of scatter, the bolometric luminosity typically falls below the peak by a factor $\sim 10$ and the lightcurve levels out \citep{vanVelzen+2019a,vanVelzen+2021,Yao+2023}.The late-time radiated energy is therefore comparable to the prompt radiated energy.

\subsection{X-rays}

Although there are a few dozen examples of TDEs whose peak X-ray luminosities are comparable to the peak optical/UV luminosities often seen, relatively few TDEs discovered by optical/UV flaring have been associated with X-ray flares, and when there are detectable X-rays during the brightest part of the flare, they are generally much less luminous than the optical/UV \citep{SaxtonKomossa2021,Hammerstein+2023}.  Nonetheless, the great majority of TDEs seen to produce X-rays, whether optically-discovered or not, have rather soft spectra during the flare: when fitted by a Planck function, the characteristic temperature $kT \sim 50 - 100$~eV \citep{SaxtonKomossa2021}.  There is, however, a good deal of diversity in their lightcurves:
in terms of how rapidly the X-rays decline \citep{Auchettl2017}, non-monotonic behavior, and separation in time from any associated optical/UV flare \citep{Jonker+2020,Malyali+2024,Wevers+2021,Wevers+2024} or spectral changes \citep{Jonker+2020,Sazonov+2020,Khorunzhev2022}.

\subsection{Radio}
Radio emission has been observed both during the prompt phase and also at times months or even years after a TDE, sometimes even when there was no prompt emission \citep{Alexander2020,Horesh2021,Cendes2023}. The luminosity of the observed radio signal is always much less than in the optical/UV or X-rays.
In most cases, equipartition analysis 
indicates that their sources involve only a small fraction of the total energy\footnote{When the TDE is jetted, as for example Swift~J1644+57,  equipartition analysis \citep{Zauderer2011,Berger2012,Barniol2013,Zauderer2013,Eftekhari2018,Yalinewich2019} suggests that the energy of the emitting electrons is comparable to the energy content of the source producing other signals (e.g. the prompt X-rays).} in the system \citep
{Krolik+2016,Alexander2016}. 
It is generally thought that the radio emission takes place at large distances from the black hole, but there is considerable controversy over the nature of the outflow producing it
\citep{Krolik+2016,Alexander2016,vanVelzen2016,Matsumoto2021}.  Because the radio source is dynamically decoupled from the bound material, we will not discuss it here.

\section{basic quantities}
\label{sec:quants}

Through dimensional analysis and simple physical arguments, the fundamental parameters of tidal disruption events determine a set of characteristic distances, timescales, and orbital properties for the debris as it separates from the star. Although these undergird the topic, it is important to recognize that details omitted from their definitions significantly alter some of the scalings they imply.

\subsection{Lengthscales}

At the most basic level, TDEs depend on the mass $\Ms$ of the star (generally stated in units of $M_\odot$), the mass of the black hole, $M_{\rm BH} $ (for which we adopt a fiducial value of $10^6 M_\odot$), and the pericenter of the star's orbit.  Because there are very few stars whose orbital semimajor axes are comparable to the scale on which TDEs happen, the rate of TDEs is almost certainly dominated by stars whose orbits are extremely eccentric.  In fact, most stars that become victims of TDEs are on effectively parabolic orbits; we will quantify this statement momentarily.

At the order of magnitude level, the criterion for the tidal gravity of the black hole to overwhelm the self-gravity of a star is
\begin{equation}
{GM_{\rm BH} R_* \over r_{\rm t}^3} \gtrsim {G\Ms \over R_*^2}.
\end{equation}
Thus, the order-of-magnitude tidal radius $r_{\rm t}$ is given by
\begin{equation}
r_{\rm t} = {R_* M_{\rm BH}^{1/3}\over \Ms^{1/3} } 
= 7 \times 10^{12} m_{\rm BH,6}^{1/3} \ms^{0.55}\hbox{~cm} 
= 50 m_{\rm BH,6}^{-2/3} \ms^{0.55} r_{\rm g} ,
\end{equation}
where $\ms$ is the stellar mass in solar masses,  $m_{\rm BH,6}$ is the black hole mass in units of $10^6 M_\odot$, and $r_{\rm g}  \equiv GM_{\rm BH}/c^2$ is the black hole's gravitational radius.
Here and in the rest of the text, we describe the main-sequence mass-radius relation by the power-law $R_* =  0.93 R_\odot \ms^{0.88}$ \citep{Ryu+2020a}. 
Whenever $\ms$ appears in a scaling relation, if its exponent is written as a decimal quantity, part of the $\ms$-dependence is through $R_*(\ms)$.

The tidal radius estimator $r_{\rm t}$  is based on an order-of-magnitude argument.  Other radii relevant to tidal disruptions differ from it by factors of order unity.  In particular, the critical radius within which a star can be completely disrupted is ${\cal R}_T = \Psi(M_\star,M_{{\rm BH}}) r_{\rm t}$ \citep{Ryu+2020a,Ryu+2020b}.
The correction factor
$\Psi(M_\star,M_{{\rm BH}})$ is defined in the Appendix.  Although $\Psi(M_\star,M_{{\rm BH}})$ is of order unity, its dependence on the stellar mass and the black hole mass is important when estimating TDE rates.

\subsection{Energy scales and orbital properties}
\label{sec:energyscales}

The specific orbital energy of the debris liberated from the star is conventionally estimated \citep{Rees1988} by
\begin{eqnarray}
\Delta E_0 &=& {GM_{\rm BH} R_* \over r_{\rm t}^2} 
= {G (M_{\rm BH} \Ms^2)^{1/3} \over R_*}  \nonumber \\
&=&\frac{GM_{\rm BH}}{r_{\rm t}} \big(\frac{\Ms}{M_{\rm BH}}\big)^{1/3}  = 2.3 \times 10^{-4}\, c^2 m_{\rm BH,6}^{1/3}  \ms^{-0.213}.
\end{eqnarray}
In other words, the orbital energy of the debris is, for typical parameters, $\sim 10^{-2}$ the potential energy near the nominal tidal radius.  However, once again, detailed calculations \citep{Ryu+2020a,Ryu+2020b,Law-Smith+2020} have found that consideration of the internal density profile of main sequence stars introduces order-unity correction factors: \citet{Ryu+2020a} defined them by $\Xi = \Delta E/\Delta E_0$, where the energy range $-\Delta E < E < +\Delta E$ contains 90\% of the debris mass.  Details about the correction factor $\Xi$, including fitting formul\ae \ for its dependence on $\Ms$ and $\MBH$, are provided in the Appendix ~\ref{App:factors}.

For the star's orbit to be ``effectively parabolic", its specific binding energy should be much smaller than the debris specific energy, $\Delta E$. This condition corresponds to an initial stellar semi-major axis much larger than the  debris minimal semi-major axis $a_0$ (defined in eqn.~\ref{eq:a0}),  and to eccentricity much closer to unity than the debris' eccentricity (defined in eqn.~\ref{eq:ecc}).

The specific orbital energy also determines another characteristic lengthscale: the semimajor axis of the initial orbit traveled by the ``most bound" matter:
\begin{eqnarray}\label{eq:a0}
    &&a_0 = {GM_{\rm BH} \over 2 \Delta E} =  \frac{M_{\rm BH}^{1/3} r_{\rm t}}{2 \Xi \Ms^{1/3} } = 
    \frac{M_{\rm BH}^{2/3} R_*}{2 \Xi \Ms^{2/3} } \\
    &&= 3.26 \times 10^{14}\, \Xi^{-1} m_{\rm BH,6}^{2/3} \ms^{0.213} \hbox{~cm}\nonumber\\
    &&=2200 \, \Xi^{-1} m_{\rm BH,6}^{-1/3} \ms^{0.22} \rg . \nonumber
\end{eqnarray}
Consistent with the ratio $\sim 10^{-2}$ between $\Delta E$ and the gravitational potential near $r_{\rm t}$, the orbital semimajor axis is $\sim 10^2 r_{\rm t}$.
Because the pericenter of all the debris orbits is very nearly the stellar center-of-mass pericenter, and $r_{\rm p}$ is often $\lesssim r_{\rm t}$, the large ratio between $a_0$ and $r_{\rm t}$ immediately implies that the debris orbits are extremely eccentric:
\begin{eqnarray}
1-e &\leq& 2(r_{\rm p}/r_{\rm t}) \, \Xi (\Ms/M_{\rm BH})^{1/3}
\nonumber\\
&\leq& 0.02 (r_{\rm p}/r_{\rm t}) \, \Xi (\ms/m_{\rm BH,6})^{1/3}.
\label{eq:ecc}
\end{eqnarray}

Lastly, the energy also determines a characteristic timescale $t_0$, the orbital period of the most bound matter:
\begin{equation} \label{eq:t0}
t_0 
={\pi \over \sqrt{2}}   \left({ M_{\rm  BH} R_*^3 \over G\Ms^2 \Xi^3 }\right)^{1/2}
=37 \, \Xi^{-3/2} m_{\rm BH,6}^{1/2} \ms^{0.32} \hbox{~d}.
\end{equation}
As $\ms$ increases, the internal density profile of main sequence stars becomes increasingly centrally-concentrated.  This causes $\Xi$ to rise as a function of $\Ms$,
and the net result is for there to be almost no net trend in $t_0$ as $\Ms$ increases from $\simeq 0.1 M_\odot$ to $\simeq 10M_\odot$ \citep{Ryu+2020a}.
On the other hand, the explicit scaling with $M_{\rm BH}$ is augmented by the implicit dependence on $M_{\rm BH}$ through $\Xi$, making $t_0 \propto \mbh6^{0.6}$.

This characteristic timescale is significant for (at least) two reasons.  The first is that it is the characteristic timescale on which the debris revisits the vicinity of the black hole: the rate at which mass returns to near the stellar pericenter rises to a peak that occurs $\simeq t_0$ after the star's pericenter passage and then declines thereafter as $(\Ms/3t_0) (t/t_0)^{-5/3}$.   The second is that, as the orbital period of the debris, it also defines the growth time of the internal stresses capable of driving accretion through outward angular momentum transport.  Nonlinear saturation of the MHD turbulence driven by the magnetorotational instability is generally thought to take $\sim 10$ orbital periods, which in this instance is $\sim 10 t_0$, or $\sim 1$~yr for typical parameters. 

\subsection{Circularization and the inverse energy crisis}
\label{sec:circ}
The length scale where most of the mass is located dictates the implied efficiency of energy extraction. The conversion efficiency of kinetic to thermal energy at distance $r$ from a black hole is:
\begin{equation}
\eta(r) 
\simeq \frac{G M_{\rm BH} }{ r c^2 } = \frac{r_{\rm g}}{r} . 
\end{equation}
For example, if, upon its first return to the black hole, the debris joins a compact disk with circular orbits of radius $\sim r_{\rm p}$, the peak rate at which orbital energy is dissipated is
\begin{equation}
L_0 \sim \eta(r_{\rm p}) \frac{ M_\star}{3t_0}
\sim  10^{46} ~ (\frac{25 r_{\rm g}}{r_{\rm p}})  \ms^{0.68} m_{\rm BH,6}^{-1/2} \Sigma^{3/2}\hbox{~erg~s$^{-1}$} . 
\end{equation}
This is $\sim 10^2 \times$ larger than the luminosity typically observed in the prompt phase, a problem that has been called the ``inverse energy crisis" (first discussed in \citet{Piran+2015}).

If this much light were generated over a surface comparable to that subtended by the inner regions of an accretion disk around a black hole, it would have a characteristic temperature
\begin{equation}\label{eq:temp_rp}
T_0 \sim \left({L_0 \over \sigma_{\rm SB} 2 \pi r_{\rm p}^2}\right)^{1/4}
\sim 1 \times 10^6 ~\left(\frac{r_{\rm p}}{25 r_{\rm g} }\right)^{-3/4} \ms^{1/6} m_{\rm BH,6}^{-5/8} \Sigma^{3/8} \hbox{~K},
\end{equation}
where $\sigma_{\rm SB}$ is the Stefan-Boltzmann constant.
This is $\sim 50 \times$ the observed optical/UV temperature \citep{vanVelzen2020SSRV,Hammerstein+2023}.
Typical observed velocities from motion at a few tens of $r_{\rm g} $  would be $\sim 30,000$~km~s$^{-1}$, about ten times larger than implied by the measured emission line widths.

Interestingly, all three  problems, about the luminosity, the temperature, and the velocity, are solved if the apocenter, $a_0$, were to replace the pericenter, $r_{\rm p} $, as the place where kinetic energy of the flow is converted to heat.

\section{Summary of current numerical simulations}
\label{sec:simulations}
 
\subsection{Realistic, Global Simulations}

Explicit hydrodynamic simulations of what happens to the bound portion of the tidal debris are the best path toward understanding the fate of the returning matter stream and uncovering the observational implications of debris dynamics.
Beginning with \cite{NoltheniusKatz1982}, many numerical hydrodynamics simulations about various aspects of TDE evolution have been published.  Unfortunately, relatively few had realistic initial conditions and were carried out long enough to be suitable for investigating the system as a whole.  The prerequisites for a simulation to be genuinely realistic and global are:
\begin{itemize}
    \item have a problem volume large enough to contain all the bound debris (or at least that portion of it returning within a few $t_0$ of the disruption);
    \item run for a time $\gtrsim t_0$;
   \item consider a star on an effectively parabolic orbit (when this criterion is not satisfied, the orbit of the returning stream is dominated by the energy of the initial stellar orbit rather than by the disruption event);
    \item self-consistently link the disruption itself with the post-disruption debris hydrodynamics;
    \item assume parameters that might apply to observed events.
\end{itemize}

Three simulations stand out as satisfying all these criteria.  We list them in order of their publication dates.
\citet{Ryu2023b} employed two fixed grids, one a small box following the star's orbit, the other a large volume surrounding the black hole. Simultaneous evolution on the two grids was coordinated through a ``multipatch" system \citep{Shiokawa+2018,Avara+2024}.   The programs running on both grids solve the general relativistic hydrodynamics equations; the one responsible for the star adds relativistically-consistent stellar self-gravity. \cite{Ryu2023b}  considered a $3 M_\odot$ star, whose internal structure was taken from a MESA model of a middle-aged main-sequence star, and a $10^5 M_\odot$ black hole.  The equation of state included LTE radiation pressure, but there was no computation of radiation transfer because, for these parameters, the cooling time was always very long compared to the evolution time.  The duration of this simulation was $3t_0$.
\citet{SteinbergStone2024} solved the equations of Newtonian hydrodynamics and radiation transport in the flux-limited diffusion approximation on a moving mesh. Stellar self-gravity was computed via a quadrupole moment tree, and the black hole's gravity was approximated as a Paczynski-Wiita potential with a softening term. They chose a $1 M_\odot$ star and a $10^6 M_\odot$ black hole, but described the internal structure of the star as an $n=3/2$ polytrope, i.e., an isentropic structure for a gas with an adiabatic index of 5/3.  This simulation ran for $\approx 1.4 t_0$.
The third \citep{Price+2024} was produced by an SPH code with relativistic hydrodynamics in a Schwarzschild spacetime and the same equation of state as in \citet{Ryu2023b}.  Their star's initial structure was the same as in \citet{SteinbergStone2024}, but its self-gravity was calculated as an integral over the Newtonian Green's function for the Poisson Equation with a softening term to smooth short lengthscale fluctuations.  The simulation covered a time up to $\simeq 9t_0$.
All three of these simulations provided resolution capable of describing the smallest relevant structures, whether in the disrupting star or in shocks within the debris flow.  \citet{SteinbergStone2024} presented an especially detailed discussion of these issues, but the strong agreement in results between the three suggests that the other two were comparably well-resolved.

Three other simulations are also of interest, but each fails one or two of the criteria.   The first \citep{Shiokawa+2015} passed all the criteria but the final one: they investigated the disruption of a $0.64 M_\odot$ white dwarf by a $500M_\odot$ black hole, choosing these parameters because they reduced the contrast in lengthscales, thereby diminishing the computational cost. Like the work of \citet{Ryu2023b}, this simulation treated the problem in terms of fully general relativistic dynamics and included radiation pressure through its LTE contribution to internal energy. It also had the longest duration of all simulations published to date: $13t_0$. The second \citep{Sadowski+2016} combined general relativistic hydrodynamics with Newtonian stellar self-gravity.  Its parameters, however, were somewhat special: the star began on a bound orbit with $e=0.97$, so that its specific orbital energy was actually greater in magnitude than $\Delta E$ of the debris.  In addition, the star's pericenter was at $7r_{\rm g} $, so this simulation probed the relatively small phase space associated with strong apsidal precession.  Lastly, the duration of the simulation was only $\simeq 0.3t_0$. Third, \citet{Andalman+2022}, much like \citet{Sadowski+2016}, used Newtonian SPH to generate tidal debris and then general relativistic hydrodynamics to study its further motions for an event in which the stellar pericenter was $7r_{\rm g}$, but the star was on an effectively parabolic orbit.  The duration was only $\approx 0.03t_0$.  Thus, it, too, cannot be directly applied to the most common events, both because it treated a TDE with an exceptionally small pericenter and because it ran for only a brief time.

\subsection{Consensus dynamical results}
\label{sec:dynamics}

All four of the simulations studying disruptions with pericenters $> 10r_{\rm g} $ found the same principal dynamical features.  The most prominent of these are several quasi-standing shocks.
As predicted by \citet{EvansKochanek1989} and \citet{Kochanek+1994}, the convergence of debris streams whose orbital planes are slightly different creates a shock near the pericenter, dubbed the ``nozzle shock".  Compared to the orbital speed, this is a weak shock, with speed at early times only $\sim (\Ms/M_{\rm BH})^{1/3} v_{\rm orb}$ and therefore able to dissipate only $\sim 10^{-4} \ms^{2/3} m_{\rm BH,6}^{-2/3}$ of the orbital kinetic energy near pericenter.  Although the nozzle shock initially comprises a pair of roughly horizontal shocks, the shock fronts tilt over time.  This results in a somewhat greater dissipation efficiency, perhaps reaching, at its greatest, $\sim (\Ms/M_{\rm BH})^{1/3}$ of the pericenter-region kinetic energy \citep{Shiokawa+2015}.  A small fraction  of the matter encountering the nozzle shock is deflected inward as the shock redistributes angular momentum (\citealp{Shiokawa+2015,Ryu2023b,SteinbergStone2024}; and \citet{Price+2024} [private communication, G. Lodato]), carrying matter to smaller radii at a rate that is at most $\lesssim 0.01 \Ms/t_0$.  The inward-moving matter does not instantaneously settle into a ``normal" circular-orbit accretion disk.  Instead, the substantial eccentricity of the deflected matter's orbits leads to further shocks at radii inside pericenter \citep{Ryu2023b,SteinbergStone2024,Price+2024}.  The pericenters of the matter gaining angular momentum move outward.  When this matter returns to the pericenter region, it shocks at a larger radius.  Consequently, the radial extent of the nozzle shock gradually stretches, reaching $\sim 400 r_{\rm g} \sim 4 r_{\rm p}$ by $t \sim 3t_0$ \citep{Ryu2023b}.

\begin{figure*}
\includegraphics[width=1.0\linewidth]{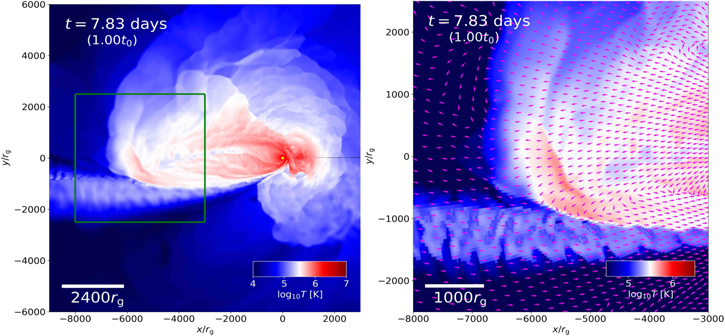}
\caption{Temperature in the orbital plane in a logarithmic colorscale (different scales in the two images), on large scales (left) and small (right) (taken from \citealt{Ryu2023b}).  The green box in the left panel shows the boundaries of the zoomed-in right panel.  Arrows in the right panel show the direction of gas flow. }
\label{fig:RyuGlobalShocks}
\end{figure*}

Debris that has passed through the nozzle shock swings back out toward an apocenter that is slightly smaller because of the orbital energy dissipated in the shock; its natal specific angular momentum is so small that the angular momentum exchange in the shock hardly affects the apocenter.  Because the disruption of the star takes place across a range of distances from the black hole (roughly from $r \simeq r_{\rm p}$ to $r \simeq 20 r_{\rm p}$: \citet{Ryu+2020b}) and because of relativistic effects in both the tidal stress and the orbits, the lines of apsides of the debris orbits stretch across a range of angles $\sim 10^\circ$ \citep{Shiokawa+2015}.   When the range of angles is this small and the orbits are highly eccentric, the streams intersect near apocenter (see also the related geometric argument of \citet{Dai+2015}).   Two views of these shocks are presented in Figure~\ref{fig:RyuGlobalShocks}, which highlights shock locations by portraying temperature (see also Fig. 1 of \citealt{SteinbergStone2024}).

Unlike the nozzle shock, the angle between the flows shocking against one another near apocenter is large.  Consequently, the energy dissipated is comparable to the local orbital kinetic energy.  In rough terms, the kinetic energy near apocenter is a fraction $1-e$ of the kinetic energy near pericenter, and we have already estimated that $1 - e \simeq 0.02 (r_{\rm p}/r_{\rm t}) \, \Xi (\ms/m_{\rm BH,6})^{1/3}$.  The energy per unit mass dissipated in an apocenter shock is then $\sim \Delta E$, comparable to or greater than that of the nozzle shock \citep{Shiokawa+2015,SteinbergStone2024}.

The precise location(s) of the apocenter shock(s) change over time.  Because the earliest matter to return is the most bound, and therefore has the smallest apocenter, at early times this shock moves outward.  At later times, after some material has gone around more than once and lost some orbital energy in the nozzle and apocenter shocks, the apocenter shock moves inward.  In addition, as the debris orbits change shape, other shocks form at radii comparable to $a_0$, but farther from the path of newly-returning debris \citep{Shiokawa+2015,Ryu2023b,SteinbergStone2024}.   Throughout these events, matter that is returning for the first time commingles with matter that has already completed one or more orbits; in other words, from very early on, the dynamics are poorly approximated by considering only a stream wrapping around the black hole once and then encountering a newly-arriving stream.   This is one of the reasons that encompassing at least the majority of the bound debris is a prerequisite for simulation credibility.

Although there have been suggestions in the literature  that the radiation pressure of light emitted from an inner accretion disk \citep{MetzgerStone2016}  or arising from ``stream-stream" shocks like the apocenter shocks \citep{LuBonnerot2020,LuBonnerot2021} could lead to unbinding a significant amount of mass, there is little evidence for this in global simulations; over the first $\sim (1 - 2) t_0$, the amount of mass converted from bound to unbound is $\lesssim 0.01 \Ms$ \citep{Ryu2023b,SteinbergStone2024}.\footnote{\citet{Price+2024} find an expanding ``Eddington envelope" carries $\sim 2/3$ of the initially bound debris outward, but do not state what fraction of this envelope is unbound.}

Lastly, despite the expectation in the traditional model that matter is accreted onto the black hole as fast as it returns from its first visit to apocenter, and therefore the possible rate of energy release is highly super-Eddington, the two global simulations with astrophysically-relevant parameters find that radiation forces are generally significant, but rarely exceed the gravitational force, i.e., the flux at most approaches the Eddington level. That this is so is particularly striking in the case treated by \cite{Ryu2023b} because the peak mass-return rate is so high that it would yield a heating rate $\sim 5000L_E$ if the efficiency were $\sim 0.1$.

\subsection{Consensus structural results}
\label{sec:structure}

As already stressed in the introduction, the simple question ``Where is the tidal debris?" is fundamental to any consideration of its observable phenomenology.  When speaking about the bulk of the bound matter that has already returned from its first visit to apocenter, the simulations give a consistent answer: at a radius $\sim a_0$.  Given the dynamical picture already summarized, it could hardly be anything else: its orbital energy has been diminished by at most a factor of order unity.  Because $a_0$ depends very weakly on both the black hole mass and the star mass (see eqn.~\ref{eq:a0}), most of the bound mass---even when it settles into an accretion flow---remains at distances $\gtrsim 2500 r_{\rm g} $ from the black hole. For example, at $t \simeq 1.4t_0$ (the endpoint of the \citet{SteinbergStone2024} simulation), $\approx 3/4$ of the accretion flow mass is $> 5000 r_{\rm g} $ from the black hole in both that simulation (private communication, E. Steinberg) and the simulation of \citet{Ryu2023b}.   At $t=3t_0$, the endpoint of the latter simulation, that fraction has hardly changed.

The vertical mass distribution for all the bound mass that has passed through at least one shock is also simply described: it is geometrically thick.  As already estimated, the apocenter shocks generically dissipate an energy similar to the net energy of the debris orbits, $\Delta E$.  It immediately follows that in energetic terms, the gas is supported as much by pressure (largely radiation) as by rotation.

The characteristic radial scale remains close to $a_0$, but the characteristic eccentricity changes substantially, dropping to $\sim 0.5$.  This is due primarily to the orbital energy lost by the gas in all of the shocks, but is aided by the angular momentum gained when the nozzle shock removes angular momentum from the fluid deflected inward and transfers it to the fluid that remains on orbits going out to the $r \sim a_0$ region.

Although the great majority of the bound debris stays relatively far from the black hole, small amounts can find their way much closer.  Again, the three extant simulations give similar results: at $t \simeq 1.4t_0$, both \citet{Ryu2023b} and \citet{SteinbergStone2024} find $\sim 2 - 5 \times 10^{-3} \Ms$ at radii $\lesssim r_{\rm p}/2$.  Both numbers, as well as the fate of this matter, are uncertain due to the inner cut-offs employed by these simulations (see the next subsection for details). Although \citet{Price+2024} do not quote a figure for the mass close to the black hole at $t \sim t_0$, their Figure~2 shows only a very small fraction of the total mass this close even at $t \simeq 9t_0$.

\subsection{Limitations}\label{sec:limitations}

To close this section, we acknowledge three limitations to the guidance we can derive from the existing global simulations. First,  two of them impose unrealistic dynamics near the black hole.  \citet{Ryu2023b} place an outflow boundary condition at a spherical radius $40r_{\rm g} $ from the black hole's center.  This means they cannot say what happens to matter passing through that surface, and they likely overestimate the net rate at which matter crosses the boundary.  Rather than place  a sharp boundary around the black hole, \citet{SteinbergStone2024} instead force the gravitational potential to increase linearly with radius from the black hole out to $30r_{\rm g} $, where it switches to a Paczynski-Wiita form.  The impact of this policy is to make the dynamics of gas close to the black hole unphysical: fluid elements have the wrong velocity and follow incorrect orbits.  The motivation for both policies is to avoid excessively short timesteps. Like \citet{Ryu2023b}, \citet{Price+2024} employ general relativistic hydrodynamics (except for the star's self-gravity), but differ by not requiring any central cut-out.  Thus, this simulation may be more reliable than the other two for matter passing close to the black hole.

The second limitation is crude approximations to time-dependent radiation transfer. In the simulations by \citet{Ryu2023b}  and \citet{Price+2024}, there is no transfer at all: the radiation is assumed to be in LTE everywhere and at all times.  Because the parameters of the \citet{Ryu2023b} simulation led to a particularly large ratio of cooling time to evolution time, this was not a bad approximation, but it would break down for larger $M_{\rm BH}$ and/or smaller $\Ms$. 
Although \citet{Price+2024} allowed no radiation transport during their hydrodynamical simulation, they estimated the radiated luminosity in post-processing.
To do so, they solved 1D time-steady transfer equations (even though the photon diffusion time was often longer than the evolution timescale) in which they assumed the total opacity had the Thomson scattering value, but was purely absorptive.  This procedure likely overestimates the luminosity.
\citet{SteinbergStone2024} solved a time-dependent radiation transfer equation along with the hydrodynamic equations, but in a very simplified form: in the gray flux-limited diffusion approximation, with Thomson scattering opacity.  In much of the volume occupied by debris, the density is low enough to support the approximation of Thomson opacity, but in the denser regions it is likely to underestimate significantly the Rosseland mean \citep{Hirose+2014}.  Moreover, the flux-limited diffusion formalism always directs the flux along the radiation intensity gradient, but this is frequently wrong near and outside the photosphere.  Errors of this sort are particularly problematic for radiation-driven outflows and geometrically-complicated photospheres.  Given these uncertainties, we note that our estimates below of the energy dissipation rate are more robust than those of the luminosity and temperature.  Nonetheless, for the angle-integrated bolometric luminosity, the time-dependent solution of \citet{SteinbergStone2024} should provide the best estimate of the three; unfortunately, it ran for only $\simeq 1.4 t_0$.

The third limitation is that all three simulations considered events with $r_{\rm p}  > 10r_{\rm g} $.  When the pericenter is smaller, relativistic apsidal precession in a single pericenter passage is a radian or more, so that strong shocks take place close to the pericenter scale, and a great deal more heat is dissipated than when the apsidal precession is weaker and the shocks take place much farther away.  This contrast can qualitatively change the character of the event, especially when $r \lesssim 6r_{\rm g} $ \citep{Ryu+2023}. Pericenter passages closer than $\sim 10r_{\rm g} $ occur in only a small fraction of disruptions by black holes with mass $\lesssim 10^7 M_\odot$, but in nearly all flare-producing events with $M_{\rm BH} \gtrsim 10^7 M_\odot$ \citep{Ryu+2020d,Krolik+2020}. For this reason, the consensus results we focus on here may need significant revision when $M_{\rm BH} \gtrsim 10^7 M_\odot$, and our statements about scaling relations may fail for $M_{\rm BH} \gtrsim  10^7 M_\odot$.

\subsection{Summary of simulation results}

For many purposes, these results can be summarized very concisely by two key statements.  It is worth remarking that their import is in excellent agreement with the results of the simulation reported by \citet{Shiokawa+2015}.

First, by a few $t_0$, when the majority of the bound mass has returned from its first passage through apocenter, a new structure has been created that is hot, irregular, crudely elliptical, and geometrically thick.  It might be called an ``accretion flow", but it is very different from a classical ``accretion disk".  During its formation, the orbits of fluid elements change by angular momentum exchange and energy dissipation in shocks, but once the accretion flow is formed, these shocks diminish in strength.

Second, that $\gtrsim 99\%$ of the debris mass remains $\sim 10^3 - 10^4 r_{\rm g} $ from the black hole at times a few $t_0$ after disruption immediately implies that ``circularization", in which debris mass is placed on roughly circular orbits at radii $\sim 2r_{\rm p}$ from the black hole, can be accomplished on a timescale $\sim t_0$ for only a very small fraction of the mass.  The accretion rate onto the black hole must therefore be at least two orders of magnitude smaller than the mass fallback rate, in sharp contradiction to the traditional model.  
This fact follows from the lack of any mechanism to remove sufficient energy from the debris; both the dissipation of orbital energy into heat and the radiation of heat by photons are far too slow (see also Sec.~\ref{sec:cooling}) to permit all but a small fraction of the debris mass to settle quickly (on the $\sim t_0$ timescale) into orbits near the black hole.\footnote{{\citet{SteinbergStone2024} extrapolated from the data of their simulation to argue that the total heating by shocks grows exponentially after $t \simeq t_0$.  However, their simulation did not extend long enough to test this extrapolation, and it is not seen in either of the two longer simulations, those of \cite{Ryu2023b} and \cite{Price+2024}.}}

\vskip 0.5cm
\section{Implications}
\label{sec:implications}

The shared results of these simulations regarding the location of the debris have many implications for our understanding of TDEs, both in terms of conceptual framework and specific observational predictions.  Remarkably, the simple fact of where the debris mass is deposited leads to strong predictions about many aspects of TDE phenomenology. Moreover, as we are about to show, these predictions are, without any fine-tuning, in agreement with many of the most striking features of these events.

\subsection{The energy budget}

As commented previously (Sec.~\ref{sec:circ}), the distance between the black hole and the debris is directly connected to the energy available to radiate. With nearly all the debris remaining (at times a few $t_0$ after the disruption) at a distance $\sim a_0$, the relevant dissipative efficiency is
\begin{equation}
\eta(a_0) 
\simeq {G M_{\rm BH} \over a_0 c^2 } \simeq \frac{r_{\rm g}}{a_0}
\simeq
4.5 \times 10^{-4} \Xi m_{\rm BH,6}^{1/3} \ms^{-0.21} . 
\end{equation}
The total heat released is then
\begin{equation}\label{eq:energyrad}
E_{\rm diss} \simeq \eta(a_0) (\Ms/2)c^2 \simeq  4 \times 10^{50}\Xi m_{\rm BH,6}^{1/3} \ms^{0.79}\hbox{~erg},
\end{equation}
right in the middle of the observed radiated energy distribution.  In fact, the total dissipated energy found by the simulations is closely consistent with this estimate.

It is possible that the $\lesssim 1\%$ of $\Ms$ deflected inside $r_{\rm p}$ adds to the dissipated energy, but as we will demonstrate in Sec.~\ref{sec:deflect}, in terms of bolometric luminosity, it can at most augment the light from the bulk of the debris by a factor of order unity.  In other words, the three simulations support the suggestion at the end of Sec.~\ref{sec:circ} about placing the debris at a distance $\sim a_0$ rather than $\sim r_{\rm p}$.

\subsection{Emission line widths}

Another consequence of our more robust knowledge of the debris' location is that it implies an orbital speed for the debris, and therefore the width of any atomic line features in its spectrum.   Although the orbital speed varies around an elliptical orbit, the scale of the speed is nonetheless determined by its semimajor axis:
\begin{equation}
v_{\rm orb} \simeq c (r_{rm g} /a_0)^{1/2} = 6400 \, \Xi^{1/2} \mbh6^{1/6} \ms^{-0.11}\hbox{~km~s$^{-1}$}
\end{equation}
This characteristic speed depends extremely weakly on both $M_{\rm BH}$ and $M_*$.  It is in the center of the measured H$\alpha$ FWHM line-width distribution for TDEs without Bowen NIII lines, and is about half the median FWHM when Bowen NIII lines are present \citep{Charalam+2022}, suggesting that perhaps Bowen lines are associated with events having larger $\ms$ and smaller $\mbh6$.

\subsection{Optical depth and the cooling time}\label{sec:cooling}

The optical depth $\tau$ of the debris is, of course, immediately determined by the density distribution.
Because the mass return-rate is $\propto t^{-5/3}$ after its peak, only a minority of the bound mass has returned to the vicinity of the black hole by a few $t_0$ after the disruption.  In addition, the mass that has returned spends most of its time near its orbital apocenter, $\gtrsim 2a_0$ from the black hole.  Consequently, the optical depth varies as a function of position.  For example, as shown in \citet{Ryu+2024}, only a fraction $f(a_0) \simeq 15 \%$ of the bound mass can be found within a distance $a_0$ of the black hole at $t=t_0$.  Over the span of radii where most of the mass is located, $0.3a_0 \lesssim r \approx 3a_0$, $f(r) \propto r^{1 + \epsilon}$, where $\epsilon$ is small and positive at $t=t_0$, but small and negative by $t=3t_0$.
If $\kappa$ is the gas's Rosseland mean opacity and $\kappa_{_{\rm T}}$ is the Thomson opacity, at $t=t_0$ the characteristic vertical optical depth to the midplane of a circular disk with radius $\sim a_0$ is
\begin{equation}\label{eq:tau}
\tau_0 \approx 80 [f(a_0)/0.15] \Xi^2 (\kappa/\kappa_{_{\rm T}})\mbh6^{-4/3} \ms^{0.55}.
\end{equation}

Because $a_0 \approx 3.5 \times 10^{14}\, \Xi^{-1} m_{\rm BH,6}^{2/3} \ms^{0.22}$~cm,
the basic dynamics of the bound debris automatically create a photosphere on a radial scale $\sim 10^{14.4\pm 0.5}$~cm, just as inferred from the radiating area associated with the optical/UV blackbody. However, it is important to note that, as shown by the simulations, the photosphere can be far from spherical, as the density distribution is both flattened and nonaxisymmetric \citep{Ryu+2024,SteinbergStone2024}.

In disk geometry, the photon diffusion time is given by $(\tau + 1) h/c$, where $h$ is the disk's vertical scaleheight and $\tau$ is the optical depth across a distance $h$ \citep{Piran+2015,Ryu+2020e}.  Because the radiation energy density is generally larger than the gas thermal energy, the characteristic cooling time for the debris within a distance $r$ of the black hole is then
\begin{equation}\label{eq:tcool}
t_{\rm cool} \approx  
10 \left[\frac{f(r)}{0.15}\right]\big(\frac{h}{r} \big)\big(\frac{a_0}{r}\big)\big(\frac{\kappa}{\kappa_{_{\rm T}}}\big) m_{\rm BH,6}^{-2/3} \ms^{0.77} \Xi \ {\rm days}\ . 
\end{equation}
 In units of $t_0$, it is \citep{Piran+2015,Ryu+2020e}:
\begin{eqnarray}\label{eq:tcoolratio}
\frac{t_{\rm cool}}{t_0} &\approx& f\frac{h}{r}  \frac{ G^{1/2} \kappa \Ms^{8/3}  \Xi^{5/2}} { 2^{3/2} \pi^2 M_{\rm BH}^{7/6} R_*^{5/2}}  
\nonumber \\
&\approx& 0.26\big(\frac{f}{0.15}\big)\big(\frac{h}{r}\big)\big( \frac{\kappa}{\kappa_{_{\rm T}}}\big)\Xi^{5/2} m_{\rm BH,6}^{-7/6} m_{*}^{0.47}  \ . 
\end{eqnarray}
The simulation results are consistent with this estimate.
For example, \citet{SteinbergStone2024} find that their proxy for the volume-integrated dissipation rate varies in a way not too different from the luminosity estimated by the flux-limited diffusion approximation (essentially equivalent to our $t_{\rm cool}$ estimate), but can occasionally depart from it by as much as a factor $\sim 10$.

Fortuitously, $t_{\rm cool}/t_0 \approx 1$ for our fiducial parameters.  Because, as we have already mentioned, \citet{Ryu+2024} found that $f(r) \propto r^{1 + \epsilon}$,  the characteristic cooling time inside the photosphere varies only slowly with radius, $\propto r^\epsilon$. When $\mbh6 \gtrsim 1$, $t_{\rm cool} < t_0$, so photon losses respond quickly to the heating rate throughout the debris and the luminosity matches the total heating rate. Within a time $\sim t_0$, radiation can vent enough heat to let the flow settle closer to its equatorial plane.  On the other hand, for encounters with a relatively low-mass black hole, the photon diffusion time is long relative to the heating time, making cooling inefficient.  In addition, in this slow-cooling regime, the internal transport of radiation is not in a steady-state.
In other words, over a time $\sim t_0$, the gas retains much of its heat content---i.e., its heat content evolves adiabatically---until close to the time its orbit carries it through the photosphere or it encounters another shock.

Because radiation transport in the slow-cooling regime is time-dependent, simple estimates of the luminosity, generally based on time-steady transfer, are subject to significant uncertainty.  However, in this context, a simple estimate of the cooling time of individual fluid elements combined with the adiabaticity of the slow-cooling regime suggests that the luminosity of a given fluid element is held approximately constant as it moves around an orbit \citep{Svirski+2017}.  Estimated in this way, the luminosity released on a radial scale $r$ is $L(r) \sim P_{\rm rad} r^2 c/\tau$, where $P_{\rm rad}$ is the radiation pressure.  With $P_{\rm rad} \propto r^{-4}$ when the volume of a moving fluid element is $\propto r^3$, as it is here due to the rough constancy of $h/r$, for a specific fluid element $\tau \propto r^{-2}$.  It follows that $L(r)$ is approximately independent of $r$.  Additionally, as we will show in the next subsection, the heating rate in TDEs links this luminosity to the Eddington luminosity.

\subsection{Flare luminosity, effective temperature, and flare duration: cooling fast and slow}

\subsubsection{Luminosity}

The mass distribution also determines the luminosity.  As argued in Sec.~\ref{sec:circ}, the depth of the potential well specifies the energy available; the luminosity follows from combining the total energy with the heating time (effectively $\sim t_0$) and the cooling time (as just estimated).
Rather than being the usual relativistic radiative efficiency times the mass fallback rate, $\sim \eta(10 r_{\rm g}){\dot M}_{\rm peak} c^2 $, the rate at which energy is dissipated is $L_{\rm peak,diss} \sim E_{\rm diss,peak}/t_0 \sim (r_{\rm g} /a_0) {\dot M}_{\rm peak} c^2 $.\footnote{Note that the $f(r)$ factor appearing in the optical depth is not relevant here because this quantity relates to the rate at which mass returns, not to how much resides within a given radius.}
The heating efficiency for the bulk of the debris is much smaller than the canonical relativistic accretion value because the gravitational potential where the apocenter shocks take place is shallower by a factor $\sim 10r_{\rm g} /a_0$.
Consequently, instead of the peak dissipation rate being extremely super-Eddington, $\sim 2 \times 10^{46} \Xi^{3/2} m_{\rm BH,6}^{-1/2} \ms^{0.66}$~erg~s$^{-1}$, it is 
\begin{eqnarray}\label{eq:Ldisspeak}
L_{\rm diss,peak} &\simeq& \eta(a_0) \frac{\Ms c^2}{3 t_0} \simeq 8.6 \times 10^{43} \Xi^{5/2} \mbh6^{-1/6} \ms^{0.47}\hbox{~erg~s$^{-1}$} \nonumber\\
&\simeq& 0.6 \Xi^{5/2} m_{\rm BH,6}^{-7/6} \ms^{0.47} \, L_E (M_{\rm BH}).
\end{eqnarray}
This much lower heating rate is greater than Eddington only for black holes with masses $\lesssim 5 \times 10^5 \ms^{0.38} M_\odot$, and even then it exceeds Eddington by much less than the expectation based on relativistic efficiency.

\begin{figure*}
\includegraphics[clip,trim={6cm 3cm 2.4cm 3cm},width=0.67\linewidth,angle=180]{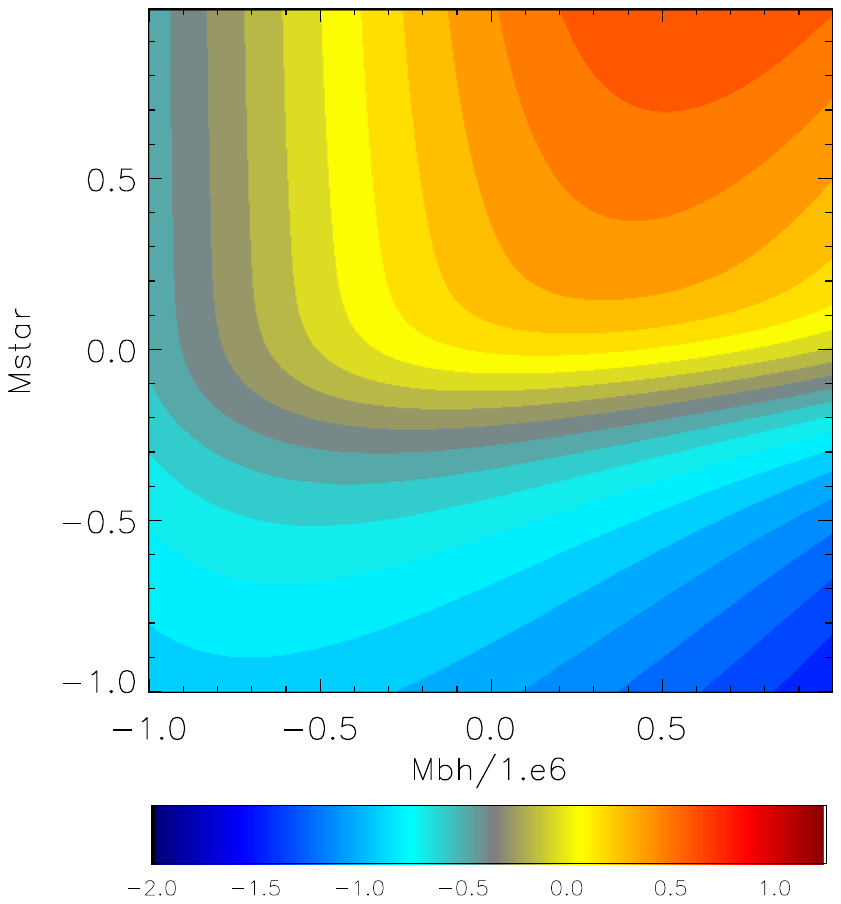} \hskip -3cm
\includegraphics[clip,trim={6cm 3cm 2.4cm 3cm},width=0.67\linewidth,angle=180]{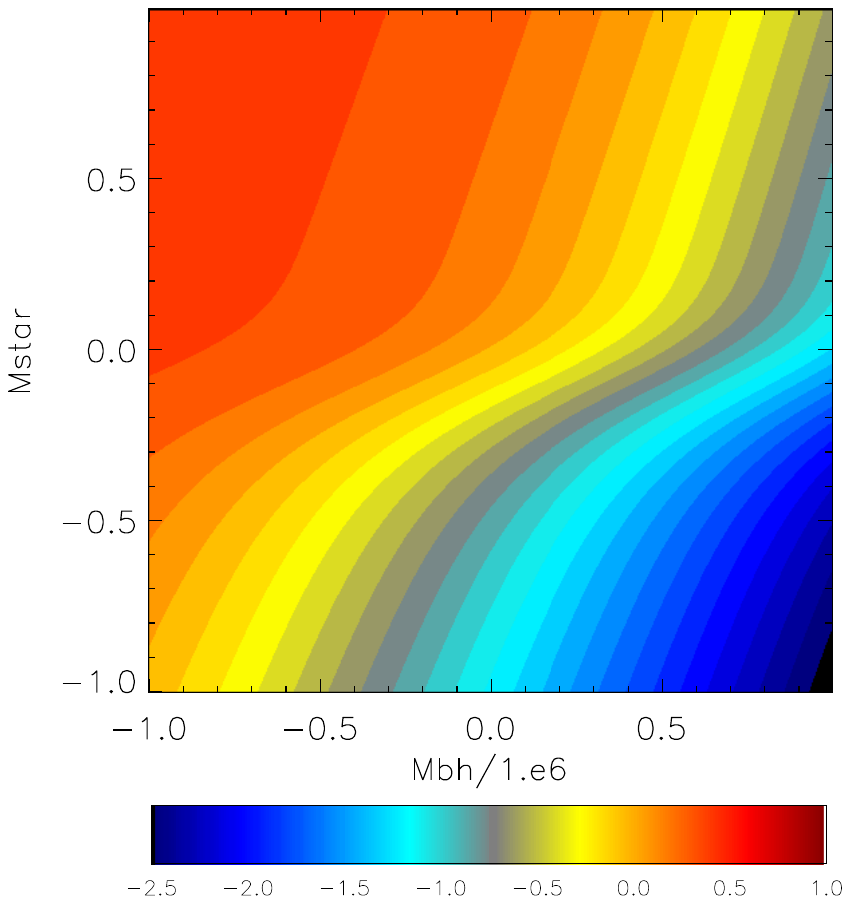}
\caption{Peak radiated luminosity as a function of $M_{\rm BH,6}$ and $\ms$ on a logarithmic scale ranging from 0.1 to 10 for both quantities.  We assume $h/r=0.3$, following the $t=t_0$ curve in Fig.~10d of \citet{Ryu+2024}. (Left) In units of $10^{44}$~erg~s$^{-1}$. (Right) In units of $L_E$. }
\label{fig:Lrad}
\end{figure*}

To translate the rate at which energy is dissipated into a radiated luminosity requires consideration of both the cooling time (see eqn.~\ref{eq:tcool})  and the duration of the heating, which is $\sim t_0$ for the bulk of the bound debris during the flare.
The radiated luminosity $L_{\rm rad} = L_{\rm diss}$ when the cooling time is shorter than the time, i.e., $t_0$, in which the heat is dissipated.  On the other hand, when $t_{\rm cool} > t_0$, the radiated luminosity is reduced by a factor $ \sim t_0/t_{\rm cool}$ relative to the heating rate $L_{\rm diss}$.
The peak luminosity (i.e., at $t=t_0$, when \citet{Ryu+2024} found the cylindrical radius of the photosphere is at $r \approx a_0$) can then be written in a combined form that approximates both the slow and fast cooling regimes:
\begin{align}\label{eq:Lradpeak}
&L_{\rm rad,peak} = {L_{\rm diss,peak} \over 1 + t_{\rm cool}/t_0}\\
&\simeq 8.6 \times 10^{43} { \Xi^{5/2} m_{\rm BH,6}^{-1/6} \ms^{0.47} \over 1 + 0.26 (f/0.15)(h/r) \Xi^{5/2} m_{\rm BH,6}^{-7/6} \ms^{0.47} (\kappa/\kappa_{_{\rm T}})}\hbox{~erg~s$^{-1}$}.
\nonumber
\end{align}
In the limit of $t_0 \ll t_{\rm cool}$, the peak luminosity is limited to roughly the Eddington luminosity and is therefore independent of $M_\star$ (see Appendix~\ref{App:Eddington}):
\begin{equation}
L_{\rm rad,peak} \simeq {2L_E \over (f/0.15) (h/r)(\kappa/\kappa_{\rm T})} \quad {\rm for}~~~ t_0 \ll t_{\rm cool}.
\end{equation}
Note that $L_{\rm rad,peak}$ may be reduced further if the density in the debris is high enough for the Rosseland mean opacity to exceed the Thomson value.  This adjustment could be important because, at the expected temperature range ($\sim 1 - 3 \times 10^4$~K), atomic features can significantly augment electron scattering when the density is $\gtrsim 10^{-10}$~gm~cm$^{-3}$ (see the opacity calculations of \citet{Hirose+2014}).

The net result for the dependence of $L_{\rm rad,peak}$ on $M_{\rm BH}$ and $\Ms$ is shown in Figure~\ref{fig:Lrad}.
In terms of the absolute flare luminosity (left panel), the greater mass of debris available from more massive stars often creates a higher radiative output.
For this reason, the highest luminosities generally correspond to the largest values of $\Ms$.
However, this no longer is the case when the larger amount of mass also means a greater optical depth and longer photon diffusion time---which occurs when the black hole mass is smaller.  This is why the peak luminosity is smaller, even for the most massive stars, when $M_{\rm BH}$ is comparatively small, and hardly depends at all on $\ms$ when $\mbh6 \lesssim 0.5$ and $\ms \gtrsim 1$.
On the other hand, larger black hole mass leads to a longer time over which radiation of nearly the same amount of energy takes place, driving down the peak luminosity for large $M_{\rm BH}$.  In the end, the highest luminosity is produced by moderate mass black holes ($M_{\rm BH} \sim 1 - 5 \times 10^6 M_\odot$) and stars of the highest mass.   The least luminosity occurs for particularly high-mass black holes and low-mass stars.  Thus, for low $M_{\rm BH}$ the optical/UV peak luminosity is $\propto M_{\rm BH}$ and almost independent of $M_\star$, while for higher black hole mass, the luminosity increases rapidly with $\Ms$ and decreases slowly with $M_{\rm BH}$.  The division between ``low" and ``high" $M_{\rm BH}$ corresponds to the peak luminosity for any particular $M_\star$, a value ranging from $\sim 1 \times 10^{43}$~erg~s$^{-1}$ for $M_\star \sim 0.1 M_\odot$ to $\sim 1 \times 10^{45}$~erg~s$^{-1}$ for $M_\star \sim 10 M_\odot$.

These predicted peak luminosities are closely consistent with the peak luminosity estimated directly from simulation data (on the basis of local cooling times by \citet{Ryu+2024}, time-dependent radiation transfer in the flux-limited diffusion approximation by \citet{SteinbergStone2024}, and instantaneous shock-heating by \citet{Price+2024}).  In all cases, the peak luminosity is predicted by our approximate expression to be $\simeq 1 \times 10^{44}$~erg~s$^{-1}$ and this coincides with the simulation-based estimates.  Note that the parameters of the disruption treated by \citet{SteinbergStone2024} and \citet{Price+2024} were the same ($M_{\star} = 1M_\odot$, $M_{\rm BH} = 1 \times 10^6 M_\odot$), whereas \citet{Ryu+2024} studied a case with $M_{\star}=3 M_\odot$ and $M_{\rm BH} = 1 \times 10^5 M_\odot$.

More importantly, for $\sim 0.3 \lesssim \ms \lesssim 3$, they are closely consistent with the peak luminosities exhibited by the bulk of the observed cases (see Sec.~\ref{sec:briefobsns} and the references cited there).
Such good agreement with the observed peak luminosities gives very strong support to this picture because its prediction of the luminosity scale has {\it no} free parameters or guessed initial conditions for the debris; it is based entirely on physical calculations beginning with a main-sequence star passing a black hole.
On the other hand, a small fraction of optical/UV TDEs shows still higher luminosities, reaching as high as $5 \times 10^{45}$ erg~s$^{-1}$ \citep{Hammerstein+2023}. As explained in Sec.~\ref{sec:rates}, very luminous events are overrepresented in a magnitude-limited sample, so their real fraction is significantly smaller than the fraction observed.
 Because the black hole masses associated with these very luminous events tend to be $\gtrsim 10^7 M_\odot$ \citep{Hammerstein+2023b,Mummery+2024,Yao+2023},
we speculate that these are the rare TDEs in which the star plunges to $r_{\rm p} \lesssim 10r_{\rm g}$, the large apsidal precession regime described at the end of Sec.~\ref{sec:limitations}.  The different dynamics of that regime may lead to their higher luminosity.

The relationship between luminosity, stellar mass, and black hole mass is simplified considerably when the luminosity is measured in Eddington units (Fig.~\ref{fig:Lrad}, right panel).  In the very long cooling time limit, the photon output is regulated to be several times $L_{\rm E}$: this regime is found for almost all $\Ms$ when the black hole mass is small.  For sufficiently high black hole mass, $L_{\rm rad,peak}/L_{\rm E}$ falls steadily but begins that fall at higher $M_{\rm BH}$ when $\Ms$ is greater.

\subsubsection{Effective temperature}

To zeroth order, we may expect a thermal spectrum for the radiation from the photosphere of the debris, although a variety of stellar-atmosphere features are likely.   Combining our estimate of $L_{\rm diss,peak}$ with our estimate of the radiating area ($2\pi a_0^2$), we arrive at a characteristic temperature  applicable to the fast-cooling limit \citep{Piran+2015,Ryu+2020e}:
\begin{eqnarray}\label{eq:Tdebris}
T_{\rm peak} &=& \left({L_{\rm rad,peak} \over 2\pi a_0^2 (1 + 2h/r) \sigma_{\rm SB}}\right)^{1/4} \\
&\simeq& 3.9 \times 10^4 \, \left[(1 + 2h/r)(1 + t_{\rm cool}/t_0)\right]^{-1/4}\Xi^{9/8} m_{\rm BH,6}^{-3/8} \hbox{~K}.\nonumber
\end{eqnarray}
Just as it did for the luminosity, a shift of radial scale for the debris from $\sim r_{\rm p}$ to $\sim a_0$ brings the temperature into the range actually observed (cf. eqn.~\ref{eq:temp_rp}). Thus, there is a strong prediction that the temperature is a few $ \times 10^4$~K, nearly independent of the star's mass or whether the debris cools rapidly or slowly---and this is exactly the temperature generally observed.
The principal parameter-dependence is on $\MBH$, which gives somewhat cooler temperatures for larger $\MBH$.

\subsubsection{Duration}

In the rapid-cooling regime, the optical/UV lightcurve should track the rate at which the debris' internal energy would increase in the absence of cooling. This rate is driven by the mass return rate, but its decline following the peak in the mass return rate is more gradual, as fluid elements continue to suffer shocks as they orbit around the black hole.  Because the placement and strength of these shocks depend on details of the flow, the degree to which the decline is slower than the decline in the mass return rate could vary from case to case, which means a small set of simulations is insufficient to predict the specific shape of the declining lightcurve.  Nonetheless, in this regime the duration of the flare should generally be a few $t_0$.

On the other hand, the peak luminosity of slowly-cooling shocked debris (i.e., the debris in events for which 
$m_{\rm BH,6 } \ll 0.3[ (f/0.15) (h/r) (\kappa/\kappa_{_{\rm T}})]^{6/7} \Xi^{15/7} m_{*}^{0.4} $)
is locked to its Eddington luminosity.  Because, in addition, the total dissipated energy available for radiation varies extremely weakly with both $M_{\rm BH}$ and $\Ms$, the optical/UV lightcurve for slowly-cooling events should have a flat peak lasting
$\simeq 0.5 (h/r)_{}(\kappa/\kappa_{_{\rm T}})\Xi \mbh6^{-2/3} \ms^{0.78}$~months, followed by a decline once the thermal energy remaining can no longer support this luminosity.  Note, however, that this limit usually applies when the black hole mass is small, making the duration typically a few months. Such a prediction was also made, for somewhat different reasons, in \citet{Metzger2022}\footnote{A number of the properties of this model overlap with ours: most of the bound debris residing on scales $\sim a_0$ for a long time, the large geometrical thickness of the flow, and, when $t_{\rm cool} > t_0$, the roughly Eddington luminosity.  The two models differ principally in that bulk kinetic energy and irregular elliptical gas flows are important in our model, but play no part in \citet{Metzger2022}.}.

Thus, the relationship between the duration of the flare peak and $t_0$ is somewhat indirect. For rapidly-cooling events, the period of high dissipation rate, and therefore maximum luminosity, lasts for $\sim t_0$ and then declines, perhaps $\propto t^{-5/3}$ and perhaps more slowly. For slowly-cooling events, the Eddington-limited phase lasts for $\simeq 0.4 \, \Xi^{5/6} \mbh6^{-1/2} m_*^{0.47}(f/0.15)(h/r)(\kappa/\kappa_T) \, t_0$.  For the smaller black holes associated with this limit, this phase has a duration $\gtrsim t_0$.

The brightest part of the flare ends once the system has cooled significantly. At this point, the accretion flow settles down to a geometrically-thin eccentric disk with a size $\sim 10^3 r_{\rm g} $ and an eccentricity $\sim 0.5$.  Further radiation during the time from $\sim 3 t_0$ to $\sim 10t_0$ comes from tapping the remaining heat in the gas or the emergence of reprocessed light initially generated by promptly-deflected matter. The latter may be significant because \citet{Price+2024} found that hydrodynamic effects led to an accretion rate onto the black hole declining rather slowly, roughly $\propto t^{-0.7}$ out to $t \sim 9t_0$.

\subsection{Scaling of flare properties with $\Ms$ and $\MBH$}

In the previous subsections we have shown how the placement of the debris mass at distances $\sim a_0$ leads to estimates of a variety of properties observable during the optical/UV flare.  All of these estimates depend, to varying degrees, on $\Ms$ and $\MBH$.   Part of the $\Ms$ and $\MBH$ dependence is derived from traditional TDE order-of-magnitude Newtonian estimates, but part comes from the results of detailed simulations informed by main-sequence internal density profiles for the stars and general relativistic effects dependent on $\MBH$.
Here we will: describe how interesting observable properties depend on the stellar and black hole mass, incorporating both kinds of dependence; and discuss how these relations can be inverted to infer $\Ms$ and $\MBH$ from observations.

\subsubsection{Total radiated energy}

The total energy dissipated during the flare $E_{\rm diss,peak} \propto \Xi \mbh6^{1/3} \ms^{0.78}$ (see eqn.~\ref{eq:energyrad}).  However, its actual scaling with $\mbh6$ is roughly $\propto \mbh6^{1/4}$ rather than $\propto \mbh6^{1/3}$ because $\Xi$ declines with increasing $\mbh6$.
On the other hand, $\Xi$ increases with $\ms$.
Consequently, the total radiated optical/UV energy during the flare's peak is a weakly rising function of $M_{\rm BH}$, but rises somewhat more rapidly with $M_\star$. 

\subsubsection{Peak optical/UV luminosity}

When cooling is rapid ($t_{\rm cool} < t_0$), equation~\ref{eq:Ldisspeak} gives the stellar and black hole mass-dependence of the emitted luminosity: $\propto \Xi^{5/2} \mbh6^{-1/6} \ms^{0.44}$.  Although the explicit dependence of $L_{\rm rad,peak}$ on $M_{\rm BH}$ is very weak, the implicit dependence through the $\Xi^{5/2}$ factor makes it a gradually declining function of black hole mass, roughly $\propto M_{\rm BH}^{-3/8}$.  On the other hand, the $\Xi^{5/2}$ factor significantly strengthens the increasing trend with $\Ms$, so that $L_{\rm rad,peak} \propto \Ms^{1.8}$ and rises by a factor $\sim 60$ from $\ms=0.1$ to $\ms=3$.
On the other hand, when cooling is slow, the peak luminosity is Eddington-limited, so $L_{\rm rad,peak} \propto \MBH$ and is independent of $\Ms$.

\subsubsection{Temperature}

As already remarked, the observed temperature is extremely insensitive to either $\Ms$ or $\MBH$.  For rapidly-cooling events, the temperature rises slightly with increasing $\Ms$ through the function $\Xi$'s dependence on the stellar mass, while it declines gradually ($\propto \mbh6^{-3/8}$) with increasing $\MBH$.  The temperature of slowly-cooling events depends extremely weakly on the masses: $\propto \Xi^{1/2} \mbh6^{-1/12} \ms^{-0.11}$.

\subsubsection{Duration}

As discussed in Sec.~\ref{sec:energyscales} and originally pointed out in \citet{Ryu+2020a} (see its Fig.~8a), the increase of $\Xi$ with $\Ms$ causes $t_0$ to be almost independent of $\Ms$.  For this reason, the timescale of a flare is a very weak indicator of $\Ms$.  It is, on the other hand, $\propto \MBH^{0.6}$.  Unfortunately, the persistence of shock heating for several $t_0$ means that the period of greatest luminosity is a few $t_0$. Quantitative prediction of the flare duration therefore demands a better calibration of ``a few" than the simulations to date can give.

\subsubsection{Parameter inference}

Parameter inference is, of course, most precise when the data's dependence on the parameter is strong.  The assembled scalings above show that the only truly strong dependence is that of the peak luminosity on $\Ms$ ($L_{\rm rad,peak} \propto \Ms^{1.8}$) when the event is in the rapid-cooling regime.

Because the radiative efficiency of the debris shocks is much smaller than the canonical relativistic radiative efficiency, events of typical luminosity can be readily explained as associated with stars with $\ms \sim 1$, rather than as very low-mass stars as has sometimes been suggested \citep{MockR-R2021,Nicholl+2022}.  In fact, because $L_{\rm rad,peak} \propto \ms^{1.8}$ when rapid-cooling applies, despite the larger numbers of low-mass stars, the frequency of low-mass star events in flux-limited samples should be rather low (see Sec.~\ref{sec:rates} and Fig.~\ref{fig:Rates}).

The next strongest dependence is the linear proportionality of $L_{\rm rad,peak}$ to $\MBH$ in the slow-cooling limit.  Regrettably, because slow-cooling applies for smaller $\MBH$, such events are harder to detect and less likely to appear in flux-limited samples.

In a previous paper on this topic \citep{TDEmass}, we presented a method (called {\sc TDEmass}) for using measurements of $L_{\rm rad,peak}$ and $T_{\rm peak}$ to infer $\Ms$ and $\MBH$.  In that work we did not include the effect of slow cooling, so we argued that $T_{\rm peak} \propto \mbh6^{-3/8}$ for all cases; allowing for slow cooling drastically weakens that dependence so that the temperature in that limit is very nearly independent of $\MBH$.

To use this method on data, we have incorporated into the existing Python code implementing {\sc TDEmass} a new algorithm to solve Eq.~\ref{eq:Lradpeak}, rather than Eq.~\ref{eq:Ldisspeak}; the code is available on github (\href{https://github.com/taehoryu/TDEmass.git}{https://github.com/taehoryu/TDEmass.git}).  We encourage interested readers to download an updated version incorporating slow-cooling effects.

\subsection{Late-time evolution}

Over longer timescales, years or more, nearly all the bound mass should eventually accrete. If most accretes with high radiative efficiency, the total energy emitted over long timescales could be as large as $\sim 10^{53}\ms$~erg, several hundred times the energy emitted during the flare.   Indeed, even observing TDEs for a few years past the flare has shown that the total energy radiated at late times, $\sim 3 \times 10^{50} \Xi^{3/2} (h/0.1r)^2 \mbh6^{-1/2} \ms^{0.18} (\Delta t/1$~yr)~erg, is comparable to the total energy radiated during the flare.   This fact, in itself, confirms what the three global simulations have shown: that the majority of the debris mass {\it cannot} have been brought to smaller radii during the flare, and that the inward progress of the debris is generically very slow. This conclusion is contrary to the assumption made in many models that the matter is deposited at small radius ($r \sim r_{\rm p} $) early on and then spreads both inward and outward \citep{Cannizzo1990,Shen2014,Mummery-SB2020}.

Much about late-time behavior is beyond what can be predicted from the three simulations' results.  For example, whether the entire $\sim 10^{53}\ms$~erg is eventually radiated is unclear.  Among numerous uncertainties, the radiated energy could be reduced to the extent that some of the gas acquires especially low angular momentum; such gas can plunge ballistically, passing through the event horizon without significant dissipation \citep{Svirski+2017}.

On the other hand, certain properties of the late-time luminosity can be discerned from what we now know.  One robust characteristic is that the inflow time is long because the mass is deposited far from the black hole. If inflow is driven by angular momentum transport, it requires internal stresses, which are generally due to correlated MHD turbulence.  Building the turbulence from scratch takes $\sim 5 - 10$ orbital periods, i.e., $\sim 5 - 10 t_0$, whether the orbits are circular or elliptical \citep{Stone+1996,HK2001,Chan+2023}.  Even after the turbulence reaches nonlinear saturation, there is considerable uncertainty about the magnitude of the resulting stresses. The inflow time is conventionally parameterized in terms of a pair of ratios, neither one easy to predict:
$t_{\rm inflow} \sim \alpha^{-1}(r/h)^2$ orbital periods, where $\alpha$ is the time- and azimuthally-averaged ratio of the vertically-integrated stress to the vertically-integrated pressure. Guessing at these parameters, one might expect \citep{Shiokawa+2015} that the timescale  on which  most of the debris accretes is $t_{\rm late} \gtrsim 2 \times 10^3 (\alpha/0.1)^{-1}(h/0.1r)^{-2} \, t_0$. Here we scale $h/r$ in terms of 0.1 because the inflow timescale is likely longer than the cooling time.
In terms of TDE parameters, this expression for the inflow time translates to
\begin{equation}
t_{\rm late} \sim 200 ~\Xi^{-3/2} (\alpha/0.1)^{-1} (h/0.1r)^{-2}m_{\rm BH,6}^{1/2} \ms^{0.82}\hbox{~yr}.
\end{equation}
We emphasize that the fiducial quantities in this expression are plausible, but hardly well-determined.  

If, as one would expect, $t_{\rm late}$ is at least several years, it immediately follows that the luminosity radiated as the matter accretes toward the black hole should not have major trends on timescales shorter than several years.   In other words, the late-time luminosity should flatten out.
In fact, the lightcurves of many TDEs do just this at times $\sim 1$~yr past peak.  When they do, the most common value of $L_{\rm late}/L_{\rm rad,peak}$ measured in terms of $\nu L_\nu$ in the NUV is $\sim 0.1$, but in some cases it is smaller by a factor of several to ten \citep{vanVelzen+2019,Yao+2023,Mummery+2024}. 
That the luminosity remains within one or two orders of magnitude of the flare peak for many years can be readily understood.  The radiative efficiency of matter accreting onto a black hole is $\sim 10^2\times$ greater than the radiative efficiency of shocks on the $r \sim a_0$ scale, and this advantage comes close to compensating for the disadvantage of a timescale $\sim 10^3\times$ longer.

Substantial scatter in the ratio of late-time to flare amplitude might be expected from a number of causes.  The saturation amplitude of the MHD turbulence may depend on accretion conditions.  With much of the debris remaining at distances $\gtrsim 10^3 r_{\rm g} $ 
while the accretion luminosity is produced, as usual, close to the black hole, some fraction of solid angle may be blocked by the debris.  The debris orbits at large distance from the black hole should align with the stellar orbit, but Lense-Thirring torques acting closer to the black hole may modulate the inflow or lead to dissipation due to mechanisms other than damping of MHD turbulence.  This list is undoubtedly incomplete.

\subsection{Matter deflected to small radii} \label{sec:deflect}

\subsubsection{Luminosity}

As mentioned in Secs.~\ref{sec:dynamics} and \ref{sec:structure}, a small amount of debris is deflected inward by the nozzle shock.  By this means, $\lesssim 1\%$ of the debris can, in fact, quickly enter an accretion flow with a radial scale comparable to the stellar pericenter.  The inflow time this close to the black hole should be much smaller than $t_0$, even after allowing for the time necessary for the MHD turbulence to reach a saturated amplitude, because the orbital period is only $\sim (\Ms/M_{\rm BH})^{1/2} t_0 \sim 10^{-3} (\ms/\mbh6)^{1/2} t_0$.  As a result, the accretion rate through this small disk should be very close to the inward-deflection rate. 
 At $t \sim 1 - 3t_0$, \citet{Ryu2023b} found the deflection rate to be $\dot M_{\rm defl} \sim 0.01\Ms/t_0$, about twice the peak black hole accretion rate found by \citet{Price+2024}, $\sim 0.005 \Ms/t_0$.
If the radiative efficiency is relativistic, the associated luminosity could be as large as
\begin{equation}\label{eq:Ldefl}
L_{\rm defl} \simeq 2.5 \times 10^{44} \left(\frac{\eta_{\rm rel}}{0.1}\right) \left(\frac{\dot M_{\rm defl}}{0.005 \ms/t_0}\right) \Xi^{3/2} m_{\rm BH,6}^{-1/2} \ms^{0.66}\hbox{~erg~s$^{-1}$}.
\end{equation}
Thus, the luminosity from the inner disk could be comparable to the luminosity from the shocked debris at much greater distance from the black hole; it would then also be mildly super-Eddington, particularly for smaller black holes and larger stars. 

Taken at face value, $L_{\rm defl}$ declines with  $M_{\rm BH}$ and increases with $M_\star$, and the explicit mass-scalings for both are steepened by the trends in $\Xi$.  However, there are numerous caveats and uncertainties attached to this estimate.   The mass deflection rate itself remains poorly determined, and its dependence on $\ms$ and $\mbh6$ is unknown.
 In addition, there are several ways the radiative efficiency of the deflected gas could be less than the canonical relativistic efficiency of 0.1.  As mentioned briefly at the end of Sec.~\ref{sec:simulations}, the debris orbits have energy much greater than that of a circular orbit with their angular momentum.  To join a small, circular orbit therefore requires the loss of a large amount of energy, and the simulations do not exhibit a mechanism for this.  If the debris orbits lose a small amount of their angular momentum, the associated gas can plunge directly across the event horizon without having lost any of its kinetic energy to dissipation \citep{Svirski+2017}.   Even if the angular momentum of the deflected debris is large enough for it to go into orbit around the black hole, it may still be small enough that Reynolds stresses suffice to push the gas onto a plunging orbit; there may be associated dissipation, but too little time to radiate \citep{Chan+2023}.  Even if there is dissipation and photons are emitted, high optical depth may suppress photon escape \citep{Price+2024} {or reprocess the radiation, as we will discuss in the next subsection.}

\subsubsection{Spectrum}
\label{sec:discuss_spectrum}

If the accreting matter close to the black hole during the flare period radiates a thermal spectrum, its temperature should be considerably higher than the light radiated by the bulk of the debris, but lower than if all the debris mass were promptly placed in such a small disk (cf. eqn.~\ref{eq:temp_rp}):
\begin{eqnarray}
T_{\rm defl} &\approx& \left({L_{\rm defl} \over 2\pi \sigma_{\rm SB} (10r_{\rm g} )^2}\right)^{1/4} 
\approx 7.5 \times 10^5 \left(\frac{\eta_{\rm rel}}{0.1}\right)^{1/4}  \times \nonumber \\ & & \left(\frac{\dot M_{\rm defl}}{0.005 \ms/t_0}\right)^{1/4} \Xi^{5/8} \mbh6^{-5/8} \ms^{0.16}\hbox{~K} \ .
\label{eq:tdefl}
\end{eqnarray}

The characteristic energy is therefore $\sim 50 - 100$~eV, i.e., soft X-rays \citep{Krolik+2016}; the systematic uncertainties expressed in the scaling factors for $\eta_{\rm rel}$ and $\dot M_{\rm defl}$ remain, but compressed by the 1/4 power relative to their influence on $L_{\rm defl}$.  Higher-mass black holes should tend to have softer X-ray spectra because,  in addition to its explicit dependence on $M_{\rm BH}$, $T_{\rm defl}$ also decreases with higher $M_{\rm BH}$ through $\Xi(M_{\rm BH})$.

It is interesting to compare this prediction to the observed spectra of steadily-accreting supermassive black holes, i.e., AGN.  For the non-blazar varieties, the spectral band contributing the most to the bolometric luminosity is the FUV \citep{Krolik1999}. That thermal emission from the inner rings produces FUV rather than soft X-rays may be understood by writing the luminosity as a fraction $\dot m$ of Eddington and supposing that it is radiated over the same area as used to estimate $T_{\rm defl}$; the result is $T_{\rm disk} \simeq 7 \times 10^5 {\dot m}^{1/4}\mbh6^{-1/4}$~K.
The lower disk temperatures in AGN can then be explained by noting that the black hole masses in AGN are often at least an order of magnitude (and sometimes several orders of magnitude) greater than $10^6 M_\odot$, and their Eddington-scaled accretion rates can be an order of magnitude below unity.

On the other hand, there is a stronger inconsistency with AGN X-ray spectra. In the typical case, the X-ray luminosity is $\sim O(10^{-1})\times$ the bolometric luminosity, and takes the form of a fairly hard power-law ($L_\epsilon \propto \epsilon^{-\alpha}$ with $0.5 \lesssim \alpha \lesssim 1$) with a high-energy cut-off at $\sim 50 - 200$~keV.  Such power-laws are only rarely seen in TDEs, and when they are, it is relatively late in the event \citep{Guolo+2024,Wen+2024} or when jets are involved \citep{Bloom+2011,Burrows+2011,Levan+2011,Cenko+2012}.  One might therefore speculate that higher $\dot m$ suppresses hard X-ray emission, a thought that receives some support from simulational studies \citep{Kinch+2021} as well as observations made of $\sim 10M_\odot$ black holes \citep{RR+McC2006}.

When we see both the FUV and the coronal X-rays from AGN, our line of sight to the central regions must be unobscured;
the UV continuum from TDEs can be seen directly because it is generated on the outside of the debris.  By contrast, soft X-rays from an inner disk in a TDE can suffer large amounts of obscuration by the surrounding debris.  The optical depth of the main body of debris is quite large: as already estimated in Equation~\ref{eq:tau}, electron scattering alone contributes an optical depth $\sim 300 \, \Xi^2 \mbh6^{-4/3}\ms^{0.55}$.  If the gas's ionization balance is near LTE, atomic absorption processes like ionization of He$^+$ should significantly add to the opacity for photons with energy $\sim 50 - 200$~eV, while K-shell photoionization of C, N, and O create substantial opacity for photons with energy $\sim 200 - 1000$~eV.  Closer to the black hole, where both the density and the temperature are higher, He$^+$ ionization contributes to the opacity wherever the temperature is $\lesssim 3 \times 10^5$~K.  Consequently, soft X-rays radiated from near the black hole would be entirely absorbed if they pass through the bound debris (in agreement with \citet{SteinbergStone2024}, who found that, averaged over solid angle, the X-ray luminosity was only $\sim 10^{-2}\times$ the optical/UV luminosity).

Note that in this sense, reprocessing of photons radiated closer to the black hole can be important.   However, it acts upon a far smaller luminosity than originally envisioned.  This sort of reprocessing also contrasts with other reprocessing scenarios in that here the gas doing the reprocessing is the bulk of the bound debris, which remains at distances from the black hole $\sim 10^3 r_{\rm g} $, rather than the debris made unbound at the time of disruption \citep{Strubbe2009} or a wind driven by super-Eddington radiation \citep{SQ2011,MetzgerStone2016}.

In addition, whether the light is scattered or absorbed and reprocessed, the large optical depth also imposes a delay between when the initial photons are radiated and when they (or the photons into which their energy is transformed) reach the photosphere. Because, as we have already shown, the photon diffusion time through the outer regions of the flow can be significant compared to $t_0$, the inner-region diffusion time can only make the total escape time longer.

Only within the small solid angle clear of obscuration would soft X-rays be observable at the time of the flare peak.   However, their consequences may be observable from a much larger solid angle because they have  energies large enough to photoionize a number of medium-$Z$ elements to unusually high ionization states.  Many of these ions have transitions linking their ground states to states only a few eV higher in energy, making the corresponding photons easily observable.
The Bowen lines frequently, but not always, seen in TDEs \citep{Blagorodnova+2019,Leloudas+2019,vanVelzen+2021} are examples: soft X-rays ionize He$^+$; recombination generates the HeII Ly$\alpha$ line; these line photons are absorbed by near-resonances in NIII and OIII (species created by UV photons below the HeII edge); collisional excitation generates optical/NUV lines from these ions.  Because the gas has much lower absorption opacity in the optical/NUV, these emission lines can (absent obscuration farther out in the galaxy) be seen even when the X-ray photons are strongly obscured. Moreover, the lengthy diffusion time for these X-rays can create delays between the peak of the optical/UV flare and the appearance of the Bowen emission lines \citep{Charalam+2022}; these delays may be augmented if the processes leading to X-ray radiation begin later than the shocks supplying the power for the optical/UV luminosity.

Depending on the ionization state of the debris gas near the edge of the optically-thin cone, reflection from the Thomson-thick debris may enhance the X-ray flux seen by observers with lines-of-sight within the cone.  Because of the several varieties of photoionization, the cone albedo is likely to have several ionization edges imprinted upon it.  At later times, when the debris has cooled, the debris opening angle widens, making higher-energy photons from near the black hole visible to a larger fraction of distant observers.  It is possible that by the time this occurs, the spectrum of X-rays radiated near the black hole may be harder.

Somewhat fortuitously (because the dynamics and initial conditions are quite different), in respect to obscuration properties the picture we have just described resembles qualitatively that put forward by \citet{Dai+2018} and \citet{Thomsen+2022}.  In the calculations reported in these papers, the gas density was assumed to be axisymmetric, to decline vertically so as to yield a constant aspect ratio $h/r = 0.3$, and to fall with increasing radius $\propto r^{-1.3}$ all the way to $r=8500 r_{\rm g} $.  The duration of the simulation, $20,000 r_{\rm g} /c\simeq 1.4 \times 10^{-6} \Xi^{3/2}\mbh6^{1/2} \ms^{-0.82} \, t_0$, was far too short for any evolution in the density at $r \gtrsim 50r_{\rm g} $.\footnote{Even though they assumed the debris contained far more magnetic flux than a star might have held, the inflow time at these radii was still far longer than the duration of their simulation.}  Consequently, the majority of the debris mass was found at $r \gtrsim 3000 r_{\rm g} $ and had an aspect ratio only slightly smaller than the one arising in the global simulations. Thus, these papers' predictions having to do with obscuration during the flare peak are similar to ours.

\subsection{Detected event rates}
\label{sec:rates}

The considerations presented so far have interesting implications for the rate of observed TDEs. Almost every astronomical survey is, in some way, flux-limited.  When the objects of interest exhibit a range of luminosities, this automatically entails a statistical bias for those with larger luminosity.  If the sources have a uniform spatial density, the number of sources in the sample is proportional to the volume out to which the sources can be detected; for non-cosmological extragalactic surveys (the case relevant to TDEs), the detected population is then $\propto L^{3/2}$.  Because the peak luminosity for TDEs is, in the end, determined by where the debris is placed after returning to the black hole, even the relative rates of events involving different stellar and black hole masses are consequences of where the debris mass goes.

\begin{figure}
\includegraphics[clip,trim={6cm 3cm 2.4cm 3cm},
width=1.3\linewidth,angle=180]
{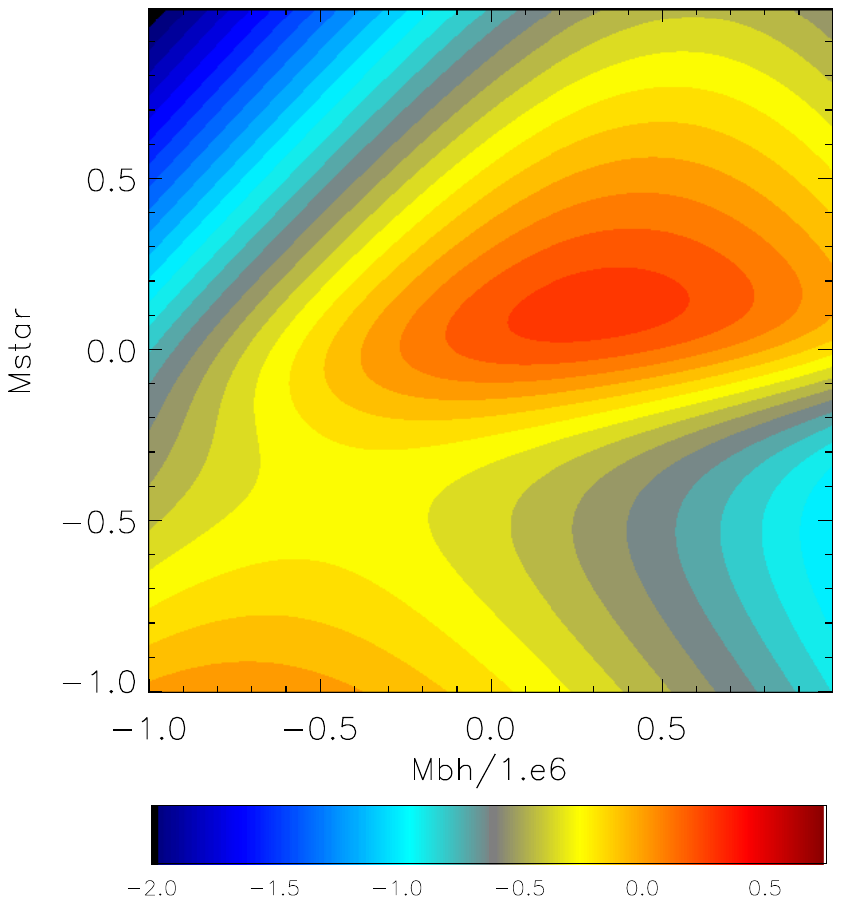}
\caption{Logarithmic contours of relative event rates $\partial^2 {\dot N}_{\rm TDE}/\partial \log M_{\rm BH}\partial \log \ms$ as found in flux-limited samples for a Salpeter mass function. The intrinsic event rate per star per galaxy is assumed to be independent of both the stellar and SMBH mass.  Both $\ms$ and $\mbh6$ are plotted logarithmically ranging from 0.1 to 10.}
\label{fig:Rates}
\end{figure}

As we have seen in Figure~\ref{fig:Lrad}, the range of predicted luminosities across the most likely portion of parameter space spans a factor $\sim 10^3$.   The most-luminous events are therefore over-represented relative to the least-luminous by a factor $\sim 3 \times 10^4$.  The sense of this over-representation is to favor events with larger $\Ms$ and intermediate $M_{\rm BH}$ (as shown in Fig.~\ref{fig:Lrad}).

In the most naive model of the intrinsic TDE population, the probability of a TDE is independent of $\ms$, so that the number of events is proportional to the stellar mass function.  Because the lifetimes of stars with $\ms \gtrsim 1$ are shorter than galaxy lifetimes, and higher mass stars yield more luminous events, the rate of events with high luminosity is very sensitive to the proportion of stars formed comparatively recently among the stars near enough the black hole to be disruption candidates.  At the other extreme, the rate of events with low luminosity is sensitive to the slope of the mass function at small $\ms$.  Here we display in Figure~\ref{fig:Rates} a simple example, a pure Salpeter mass function extending up to $10 M_\odot$, and postpone more detailed discussions of event rates to another paper.

The shape of the detected events distribution in the $M_{\rm BH} \times M_*$ plane reflects two trade-offs.  In rough terms, the dependence on $M_*$ balances the greater intrinsic number of low-mass stars against the higher luminosity of high-$M_*$ disruptions.  The dependence on $M_{\rm BH}$ peaks at intermediate masses because events at lower black hole mass are limited to $\sim L_E \propto M_{\rm BH}$ and events at higher mass have lower peak luminosities because the timescale is longer. In this example using a Salpeter mass function, the luminosity bias places the highest event detection rate at $M_\star \sim 1 - 1.5M_\odot$ and $M_{\rm BH} \sim 0.5 - 2\times 10^6 M_\odot$.  The lowest rates
are found at low stellar mass and high black hole mass ($M_\star \lesssim 0.5M_\odot$ and $M_{\rm BH} \gtrsim 5 \times 10^6M_\odot$), and at high stellar mass and very low black hole mass: $\Ms \gtrsim 5 M_\odot$ and $M_{\rm BH} \lesssim 2 \times 10^5 M_\odot$.

\section{Summary}
\label{sec:conclusions}
In the past year, three global TDE simulations in which events with astrophysically-realistic parameters were studied ($\Ms \sim 1 M_\odot$ and $\MBH \sim 10^5 - 10^6 M_\odot$) have been published.  Although they used different hydrodynamic algorithms (a fixed-grid intrinsically-conservative method, a moving-mesh intrinsically-conservative method, and SPH), in many respects their results are very consistent.   In particular, all three find that after a few characteristic orbital times for the debris, the overwhelming majority of the debris mass---$\gtrsim 99\%$---is found at radii $\sim 3 \times 10^3 r_{\rm g}  $.  Thus, this fact is now a robust prediction of the basic physics of these events, subject to alteration only for extreme parameter choices (e.g., stellar pericenters $\lesssim 10r_{\rm g} $, for which strong apsidal precession would change the dynamics \cite{Krolik+2020}.). 

Because the location of the mass directly implies many physical quantities: orbital binding energy, orbital timescale, photospheric area, etc., the robust agreement on where the mass goes leads to a similarly robust agreement in their predictions of important observable properties.  Although in all three simulations the specific parameters chosen were selected only for being within the plausible range and for computational convenience, the simulations' predictions for the principal observables associated with the optical/UV flare lie squarely in the middle of the observed distribution  \citep[as compiled, e.g., by][]{Hammerstein+2023,Yao+2023}: total energy radiated a few $\times 10^{50}$~erg, peak optical/UV luminosity $\sim 1 \times 10^{44}$~erg~s$^{-1}$, and temperature of a black-body component $\simeq 3 \times 10^4$~K.
This agreement with observed properties is especially striking because it is achieved entirely on the basis of physical calculations without any adjustment of parameters. Note, however, that our model may not be applicable to events in which $\MBH \gtrsim 10^7 M_\odot$ because in this black hole mass range the critical pericenter for a full disruption is $< 10 r_{\rm g} $, a regime in which apsidal precession is strong enough to alter the flow pattern found in the three simulations.  It is possible that the small number of events with $L_{\rm rad,peak} \gtrsim 10^{45}$~erg~s$^{-1}$ take place in this regime. 

In addition, the fact that nearly all the mass is deposited at such a large distance from the black hole directly implies that its subsequent accretion will stretch over a long period of time.   It immediately follows that after the flare, the luminosity should decline over a period of several times the characteristic debris orbital timescale until it flattens out and maintains a roughly constant luminosity for many years.
This prediction, too, is only weakly dependent on parameters and is nonetheless in excellent agreement with observations.  In fact, because the total energy radiated on several year-timescales is generically comparable to the energy radiated during the flare \citep{vanVelzen+2019a}, observations {\it demand} a situation in which accretion of a large part of the debris mass takes place slowly.

All three simulations predict similar luminosities because the underlying mechanism is also similar: shocks dissipating $\sim 10^{-3}$ of the debris' rest-mass energy on a timescale $\sim t_0$.  For this reason, all three likewise share a scaling relationship for how the luminosity depends on $\Ms$ and $\MBH$.   Because the accessible volume for flux-limited surveys is $\propto L^{3/2}$, the rate at which events are detected in such surveys is $\propto L(\Ms,\MBH)^{3/2} \partial^2 N_{\rm TDE}/\partial \Ms \partial \MBH$.  Thus, the results of these simulations predict how much events with specific pairs of $\Ms$ and $\MBH$ are favored (or disfavored) for appearance in surveys, relative to their true rate; the highest luminosities are always strongly over-represented; events with $10\times$ the typical luminosity are over-represented by a factor $\sim 30$.  Interestingly, the rate at which events are found if the stellar mass function takes the Salpeter form and there has been relatively recent star formation near the galaxy center is within a factor $\sim 3$ of its greatest amount for $0.6\lesssim \Ms/M_\odot \lesssim 6$ and $2 \times 10^5 \lesssim \MBH/M_\odot \lesssim 1 \times 10^7$. All the independent estimates of SMBH masses in the sample of \cite{TDEmass} are in this range, as are 28 of the 33 independently estimated SMBH masses in the \citet{Yao+2023} compilation.

All three simulations also agree that the fraction of the debris pushed quickly to radii within the stellar pericenter is $\lesssim 1\%$.  Even though this is a very small fraction, it may possibly contribute to the bolometric luminosity of the flare at a level comparable to the bulk of the debris; its orbital binding energy is $\sim O(10^2)\times$ that of the bulk.  This material may be the source of the soft X-rays sometimes seen in TDEs, {and reprocessing of these soft X-rays at larger distances by the bulk of the debris may lead to the Bowen emission lines often seen.}  However, none of the simulations done so far has treated the inner region with sufficient care to determine either its immediate radiation properties or, at a quantitative level, how much obscuration and reprocessing the emitted light may suffer {\it en route} to distant observers.

Thus, the basic mechanics of the most common variety of events is now understood.  Over timescales of a few $t_0$, very nearly all the debris retains its original orbital energy to within a factor $\sim O(1)$; in so doing, it remains a distance $\sim a_0$ from the black hole.   The optical/UV flare results from radiating the energy that is dissipated.   This matter approaches closer to the black hole only over much longer (years +) timescales.   A small fraction ($\lesssim 1\%)$ of the mass moves inward more rapidly.  It should also be emphasized, however, that rarer varieties of TDEs---those with pericenters small enough to create large apsidal precession \citep[e.g.,][]{Ryu+2023}, or large enough to produce a partial disruption \citep[e.g.,][]{Ryu+2020c,Liu+2024,Sharma+2024}, or those in which a massive disk already orbits the black hole \citep{Chan2019,Chan2021,Chan+2022,McKernan+2022, Ryu+2024}, for example---may behave differently.

\begin{acknowledgments}
We acknowledge support
from the National Science Foundation (NSF) grants AST-2009260 and PHY-2110339
(JHK). We also received support from the European Research Council Advanced Grant ``MultiJets" and grant MP-SCMPS-00001470 from the Simons Foundation
to the Simons Collaboration on Extreme Electrodynamics of Compact Sources - SCEECS (TP). 
\end{acknowledgments}

\begin{appendices}
\section{Appendix - Stellar Structure and Relativistic Correction Factors}
\label{App:factors}
Two correction factors are important in estimating the depends of observational properties on the stellar and the black hole masses. The first factor determine the relation between the real tidal radius ${\cal R}_T$ and the order of magnitude estimate $r_{\rm t}$:
${\cal R}_T = \Psi(M_\star,M_{_{\rm BH}}) r_{\rm t}$ \citep{Ryu+2020a,Ryu+2020b}.
\begin{eqnarray}
\Psi(\ms,\mbh6)&=&\big[ 0.80 + 0.26~\mbh6\big]^{0.5}\\ \nonumber 
&\times & \frac{1.47+ ~\exp[(\ms-0.669 )/0.137]}{1 + 2.34~\exp[(\ms-0.669)/0.137]}.
\label{eq:psi}
\end{eqnarray}
$\Psi$ is a function of $m_{*}$, and possibly the star's age and chemical composition, through the dependence of the star's internal density profile on these factors. Taken as a function of $\ms$ alone, it declines smoothly from $\approx 1.5$ for very small $\ms$ to $\approx 0.43$ for all $\ms \gtrsim 1.$. For a typical MS star encountering a black hole with $M_{\rm BH} = 10^6 M_\odot$, ${\cal R}_T \approx 25 \rg $ almost independent of the stellar mass \citep{Ryu+2020a,Ryu+2020b,Law-Smith+2020}.  This ``physical tidal radius" is also a function of $M_{\rm BH}$ through the impact of general relativistic corrections; with $r_{\rm t}$ several tens of gravitational radii, they can be substantial \citep{Ryu+2020a,Ryu+2020d}:  the portion of $\Psi$ dependent upon $M_{\rm BH}$ grows from $\approx 1.8$ for $\mbh6 \approx 10$ to $\approx 5$ for $\mbh6 = 100$.

\begin{figure}
\includegraphics[clip,trim={6cm 3cm 2.4cm 3cm},width=1.4\linewidth,angle=180]{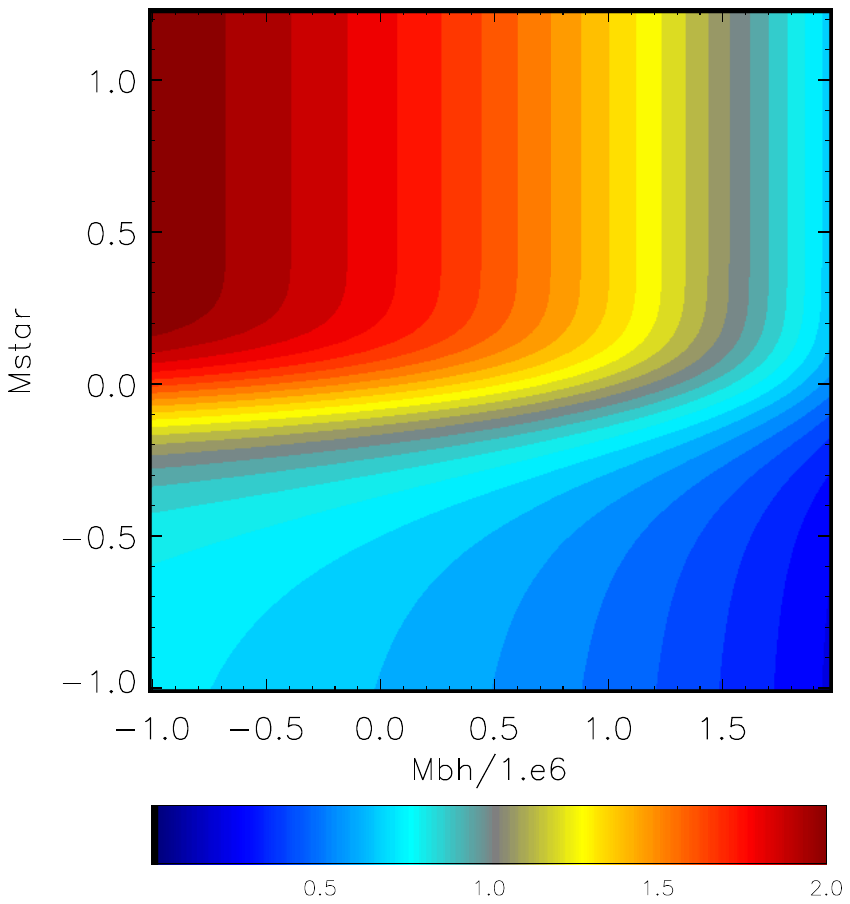}
\caption{Linear color contours of $\Xi(\ms,\mbh6)$ as a function of the logarithms of $\ms$ and $\mbh6$.  }
\label{fig:Xi}
\end{figure}

The second correction factor $\Xi(M_\star,M_{_{\rm BH}})$ defines the change in energy of the debris, $\Delta E$,  relative to the fiducial change $\Delta E_0$:
\begin{eqnarray}
\Xi(\ms,\mbh6) &\equiv& \big[1.27 - 0.3\mbh6 ^{0.242}\big]     \\ \nonumber &\times&
\frac{0.62+\exp{[(\ms-0.67)/0.21]}}{1 + 0.55~\exp{[(\ms-0.67)/0.21]}} \ . 
\end{eqnarray}
Like the factor $\Psi$, $\Xi$ is a function of stellar mass, with additional possible dependence on stellar age and chemical composition.
The function $\Xi(\ms)$ is almost opposite in behavior to $\Psi(\ms)$: $\Xi$ rises from an asymptote $\simeq 0.66$ for $\ms \lesssim 0.3M_\odot$ to an asymptote at $\approx 1.8$ for $\ms \gtrsim 1.3 M_\odot$.
$\Xi$ also depends upon $M_{\rm BH}$, particularly when $\mbh6 \gtrsim 1$, because of general relativistic effects \citep{Ryu+2020a,Ryu+2020d}:  as a function of $M_{\rm BH}$, $\Xi$ falls from $\simeq 1.3$ in the Newtonian limit to $\approx 0.75$ when $\mbh6 = 10$ and to $\approx 0.3$ at $\mbh6 = 30$, beyond which the fitting formula no longer applies. Figure~\ref{fig:Xi} illustrates how $\Xi$ is greatest for large $\ms$ and small $\mbh6$ and least for small $\ms$ and large $\mbh6$.

\section{Appendix - The Eddington luminosity limit}
\label{App:Eddington}
That the luminosity is limited to approximately Eddington is, in fact, a general result when a gas is heated to nearly the virial temperature (in the sense of associated photon energy density per unit mass), and is then left to radiatively cool in an environment whose opacity  is $\sim \kappa_T$ \citep{Krolik2010,Chan2020}.   This may be seen from a very simple argument.  If $U_{\rm rad}/\rho \sim GM /r$ in a homogeneous sphere, the luminosity is roughly
\begin{equation}
L \sim {GM M_g \over r} (c/r) \left[1 + 3\kappa_T M_g/(4\pi r^2)\right]^{-1},
\end{equation}
where $M$ is the central mass and $r$ is radius of the sphere.  The optical depth $\tau_T (r)\sim (3/4\pi)\kappa_T M_g/r^2$.  Using the definition $L_E = 4\pi c GM/\kappa_T$, this approximation to the luminosity  becomes
\begin{equation}
L \sim {L_E \over 3 (1 + 1/\tau_T(r))}.
\end{equation}
Thus, the luminosity is $\sim L_E \tau_T$ when $\tau_T < 1$ and rises to $\sim L_E$ when $\tau_T > 1$.

\end{appendices}

\bibliography{biblio}

\end{document}